\documentclass[12pt, a4paper]{article}
 \usepackage[font=small,format=plain,labelfont=bf,up,textfont=normal,up,justification=justified,singlelinecheck=false]{caption}
\usepackage{subcaption}
\usepackage{mathrsfs}
\usepackage[a4paper, left=2cm,right=2cm]{geometry}
\usepackage[colorlinks=true,linkcolor=black,citecolor=teal,urlcolor=MidnightBlue,filecolor=black]{hyperref}
\usepackage{amsfonts}
\usepackage{setspace}
\usepackage{slashed}
\usepackage{braket}
\usepackage{amsthm,amsmath,amssymb}
\usepackage{mathrsfs}
\usepackage[dvipsnames]{xcolor}
\definecolor{SchoolColor}{rgb}{0.6471, 0.1098, 0.1882} 
\usepackage{subfloat}

\usepackage[utf8,applemac]{inputenc}
\usepackage{tensor}
\usepackage{cite}
\usepackage[thicklines]{cancel}
\usepackage{tikz,tikz-cd,tkz-euclide}\usetikzlibrary{cd}\usetikzlibrary{calc} \usetikzlibrary{patterns}
\usepackage[compat=1.1.0]{tikz-feynman}
\usetikzlibrary{arrows.meta}

\usepackage{graphicx}
\usepackage{bm} 
\allowdisplaybreaks[4]
\usepackage[utf8,applemac]{inputenc}
\usepackage{tensor}
\usepackage{cite}
\usepackage{tikz}
\usepackage{graphicx}
\graphicspath{{figure/}}
\usepackage{array}
\usepackage{booktabs}

\usepackage{multirow}

\usepackage{dcolumn}
\usepackage{bm}

\usepackage{verbatim}

\usepackage{textcomp} 
\usepackage{graphicx} 

\numberwithin{equation}{section}
\newcommand{\bea}{\begin{eqnarray}}
\newcommand{\eea}{\end{eqnarray}}
\newcommand{\be}{\begin{equation}}
\newcommand{\ee}{\end{equation}}
\newcommand{\bs}{\begin{subequations}}
\newcommand{\es}{\end{subequations}}
\def\nn{\nonumber}

\def\fg{\mathfrak{g}}

\def\p{\partial}

\newcommand{\cR}{\mathcal{R}}

\newcommand{\n}{\nabla }
\newcommand{\beqs}{\begin{eqnarray}}
\newcommand{\eeqs}{\end{eqnarray}}
\numberwithin{equation}{section}

\newcommand{\Rmnum}[1]{\uppercase\expandafter{\romannumeral #1\relax}}

\def\cc{\mathcal{C}}\def\ce{\mathcal{E}}\def\cf{\mathcal{F}}\def\ci{\mathcal{I}}\def\ck{\mathcal{K}}\def\cl{\mathcal{L}}\def\cm{\mathcal{M}}\def\co{\mathcal{O}}\def\cp{\mathcal{P}}\def\cq{\mathcal{Q}}\def\ct{\mathcal{T}}

\def\by{{\bar{y}}}\def\bz{{\bar{z}}}

\def\c.c.{\mathrm{c.c.}}
\def\mn{{\mu\nu}}\def\rs{{\rho\sigma}}
\def\a{\alpha}\def\g{\gamma}

\def\z{\zeta}
\def\th{\theta}
\def\k{\kappa}\def\l{\lambda}
\def\m{\mu}\def\n{\nu}
\def\r{\rho}\def\s{\sigma}
\def\om{\omega}
\def\Th{\Theta}\def\Om{\Omega}

\newcommand{\lon}{\color{blue}}

\begin{document}
\begin{titlepage}

\begin{flushright}\vspace{-3cm}
{\small
\today }\end{flushright}
\vspace{0.5cm}
\begin{center}
	{{ \LARGE{\bf{Quantum flux operators in the fermionic theory and their supersymmetric extension\vspace{8pt}\\  }}}}\vspace{5mm}
	
	\centerline{\large{Si-Mao Guo$^{*,\dagger}$\footnote{smguo@ihep.ac.cn}, Wen-Bin Liu$^{\dagger}$\footnote{liuwenbin0036@hust.edu.cn} \& Jiang Long$^{\dagger}$\footnote{longjiang@hust.edu.cn}}}
	\vspace{2mm}
	\normalsize
	\bigskip\medskip

	\textit{* Institute of High Energy Physics, Chinese Academy of Sciences, Beijing 100049, China
	}\vspace{1mm}
	
	\textit{$\dagger$ School of Physics, Huazhong University of Science and Technology, \\ Luoyu Road 1037, Wuhan, Hubei 430074, China
	}
	
	\vspace{25mm}
	
	\begin{abstract}
		\noindent
		{We construct quantum flux operators with respect to the Poincar\'e symmetry in the massless Dirac theory at  future null infinity. An anomalous helicity flux operator  emerges from the commutator of the superrotation generators. The helicity flux operator corresponds to the local chiral symmetry which is the analog of superduality in the gauge theories. We also find its relation to the non-closure of the Lie transport of the spinor field around a loop. We discuss various algebras formed by these operators and constrain the test functions by the requirement of eliminating the non-local terms and satisfying  the Jacobi identities. Furthermore, we 
		explore their $\mathcal{N}=1$ supersymmetric extension in the Wess-Zumino model. There are  four kinds of quantum flux operators, which correspond to the supertranslation, superrotation, superduality and supersymmetry, respectively. Interestingly, besides the expected supertranslation generator, a helicity flux operator will also emerge in the commutator between the superflux operators. We check that our flux algebra can give rise to the super-BMS and super-Poincar\'e algebras with appropriate choice of parameters. In the latter reduction, we find the helicity flux reduces to behaving like a $R$ symmetry generator in the commutator with the superflux. For completion, we derive the $R$ flux which also includes a charge flux for complex scalar besides the  helicity flux for spinor field. }\end{abstract}
	

\end{center}

\end{titlepage}
\tableofcontents
\section{Introduction}
Null hypersurfaces, such as the future/past null infinity of an asymptotically flat spacetime and the event horizon of a black hole, are extremely important in many physical systems. Intrinsically, one of the amusing facts of the null hypersurface is that its metric is degenerate, which is distinguished from a Riemann manifold. Recently, this kind of null geometry has been studied in the context of Carrollian manifolds \cite{Une, Gupta1966OnAA,Duval_2014a,Duval_2014b,Duval:2014uoa}. Similar to the conventional Riemann manifold, there is a Carrollian diffeomorphism associated with any Carrollian manifold \cite{Ciambelli:2019lap}. 
The Carrollian diffeomorphism is shown to preserve the Carroll structure \cite{Liu:2022mne} and could accommodate  various extensions of the BMS group \cite{Bondi:1962px,Sachs:1962wk,Barnich:2009se,Barnich:2010eb,Campiglia:2014yka,Campiglia:2015yka,Freidel:2021fxf}. The Carrollian diffeomorphism can be  performed by quantum flux operators, which characterize the time and angle dependence of the Poincar\'e fluxes across the Carrollian manifold \cite{Liu:2022mne}.  For the spinning theories, the Carrollian diffeomorphism has been extended to intertwined Carrollian diffeomorphism \cite{Liu:2023qtr,Liu:2023gwa,Liu:2023jnc} after including the helicity flux operator.

There are plenty of interesting features for the helicity flux operator. At first, the helicity flux operator generates superduality transformation on the boundary field which extends the electromagnetic duality (and its gravitational extension) in the bulk \cite{Dirac:1931mon,Deser:1976iy,Henneaux:2004jw} to a local rotation at the boundary. The latter is isomorphic to the local rotation of the vielbein field on the celestial sphere \cite{Liu:2024llk}. 
Second, the helicity flux operator encodes the angle-dependent information of the difference between  numbers of particles with opposite helicities, which are interesting observables related to the  non-linear spin memory effect \cite{Seraj:2022qyt,Dong:2024ily,Faye:2024utu}. Third,  it has been shown that the global part of the helicity flux operator could be obtained by reducing the topological Chern-Pontryagin term to the null boundary \cite{Liu:2024rvz}.

However, there are still no systematic investigations on the helicity flux operator at future/past null infinity for  the fermionic field. We will try to fill this gap in this paper which is urgent due to the following reasons. At first, since the Chern-Pontryagin term leads to the helicity flux operator and is also related to the ABJ anomaly \cite{Adler:1969gk,Bell:1969ts}, the helicity flux operator for fermionic field is expected to concern chiral symmetry. Second, the supersymmetric BMS algebra has been discussed by \cite{Awada:1985by}, and promoted by celestial holography recently \cite{Dumitrescu:2015fej,Avery:2015iix,Henneaux:2020ekh,Fuentealba:2021xhn,Fotopoulos:2020bqj,Prabhu:2021bod,Banerjee:2022abf,Banerjee:2022lnz}. From the perspective of Carrollian holography \cite{Donnay:2022aba,Bagchi:2022emh}, it would be curious to understand how to combine superduality and supersymmetry in the framework of bulk reduction \cite{Liu:2024nkc}. 

In this paper,  we will discuss various algebras formed by the quantum flux operators. Interestingly, we can constrain the test functions in the quantum flux operators by requiring the Jacobi identities. As a consequence, the test function of the general supertranslation is at most a quadratic polynomial of the retarded time which is more restricted than our previous treatment. Besides various algebras found in the literature, we can also obtain several interesting new algebras by including the helicity flux and superflux operators.

The structure of the paper is as follows. We will explore the bulk reduction for the Dirac theory and construct quantum  flux operators in section \ref{dirac}. The helicity flux operator is found by computing  the commutators of the superrotation generators and requiring the closure of the Lie algebra. Based on the result of the Dirac theory, we extend the discussion to the Wess-Zumino theory in the next section. We will conclude in section \ref{dis}. Identities associated with twistors are collected in appendix \ref{twistor}. 

\section{Dirac theory}\label{dirac}
In this section, we will discuss on  the construction of quantum flux operators in Dirac theory where many technical treatments can be extended to the Wess-Zumino model in the next section. We will first  introduce the conventions of this work in subsection \ref{conv} and solve the Dirac equation near $\mathcal{I}^+$ to obtain the boundary radiative mode $F$ and $G$ in  subsection \ref{eom}. We will discuss the canonical quantization of the boundary fields in the following subsection. The quantum flux operators associated with Poincar\'e and chiral symmetry are constructed in subsection \ref{qfo}. We will build the connection between the quantum flux operators and bulk Lie derivative of the spinor field in  subsection \ref{stsr} and compute the flux algebra in  subsection \ref{icd}.

\subsection{Conventions}\label{conv}
As a warm up, we will reduce the Dirac theory to $\mathcal{I}^+$. We will define the Dirac spinor as 
\be 
\Psi=\left(\begin{array}{c}\chi_a\\ \xi^{\dagger \dot a}\end{array}\right)
\ee where the dotted and undotted indices denote the left and right-handed spinors, respectively
\be 
a=1,2,\quad\text{and}\quad\dot a=\dot 1,\dot 2.
\ee We will use the $\epsilon$ symbols to raise and lower spinor indices with the convention 
\be 
\chi^a=\epsilon^{ab}\chi_b,\quad \chi^\dagger_{\dot a}=\epsilon_{\dot a\dot b}\chi^{\dagger\dot b},\quad \chi_a=\epsilon_{ab}\chi^b,\quad \xi^\dagger_{\dot a}=\epsilon_{\dot a\dot b}\xi^{\dagger\dot b},
\ee where 
\bea 
\epsilon^{ab}=\epsilon^{\dot a\dot b}=\left(\begin{array}{cc}0&1\\ -1&0\end{array}\right),\quad \epsilon_{ab}=\epsilon_{\dot a\dot b}=\left(\begin{array}{cc}0&-1\\ 1&0\end{array}\right).
\eea 
Note that the right-handed spinor is always associated with a dagger  to distinguish with the left-handed spinor. The dagger can be understood as the  Hermitian conjugate
\be 
\chi^\dagger_{\dot a}=\left(\chi_a\right)^\dagger.
\ee It is convenient to omit the indices when a pair of indices are contracted as follows
\be 
\chi\eta=\chi^a\eta_a,\quad \chi^\dagger\eta^\dagger=\chi^\dagger_{\dot a}\eta^{\dagger \dot a}.
\ee 
Since the spinor fields anticommute 
\be
\chi_a\eta_b=-\eta_b\chi_a,\quad \chi^\dagger_{\dot a}\eta^{\dagger}_{\dot b}=-\eta^{\dagger}_{\dot b}\chi^\dagger_{\dot a},
\ee
we have a nice property
\be 
\chi\eta=\eta\chi,\quad \chi^\dagger\eta^\dagger=\eta^{\dagger}\chi^\dagger.
\ee 
With this convention, the Hermitian conjugate of the Dirac field is 
\be 
\Psi^\dagger=(\chi^\dagger_{\dot a},\xi^a).
\ee 
The action of a massless Dirac field reads
\bea 
S[\Psi]=\int d^4x\, i\bar\Psi \slashed\partial \Psi,
\eea where $\slashed\partial=\gamma^\mu\partial_\mu$ and the $\gamma^\mu$ matrix is 
\bea 
\gamma^\mu=\left(\begin{array}{cc}0&\sigma^\mu\\\bar\sigma^\mu&0\end{array}\right)
\eea with 
\bea 
\sigma^\mu_{a\dot a}=(1,\sigma^i),\quad \bar\sigma^{\mu\dot a a}=(1,-\sigma^i).
\eea The three Pauli matrices are chosen as 
\bea 
\sigma^1=\left(\begin{array}{cc}0&1\\1&0\end{array}\right),\quad \sigma^2=\left(\begin{array}{cc}0&-i\\ i&0\end{array}\right),\quad \sigma^3=\left(\begin{array}{cc}1&0\\0&-1\end{array}\right)
\eea such that the $\gamma^\mu$ matrices obey the anti-commuting relations
\be 
\{\gamma^\mu,\gamma^\nu\}=-2\eta^{\mu\nu},
\ee where the inverse of the Minkowski metric is 
\be \eta^{\mu\nu}=\left(\begin{array}{cccc}-1&0&0&0\\ 0&1&0&0\\ 0&0&1&0\\ 0&0&0&1\end{array}\right).
\ee In the action, we have introduced the Dirac conjugate spinor $\bar\Psi$ as follows
\be 
\bar\Psi=\Psi^\dagger \left(\begin{array}{cc}0&1\\1&0\end{array}\right)=(\xi^a,\chi^\dagger_{\dot a}).
\ee 

We now introduce the conventions on spacetime. In the retarded coordinates $(u,r,\th^A)$ with $\th^A=(\th,\phi)$ for the sphere, the flat metric reads 
\begin{align}
    ds^2=-du^2-2dudr+r^2(d\th^2+\sin^2\th d\phi^2).\label{metric}
\end{align}
The retarded coordinates are related to the Cartesian coordinates through
\be 
x^\mu=u\bar m^\m+r n^\mu,
\ee  
where $\bar m^\m=(1,0)$ which can be expressed as
\begin{align}
    \bar m^\m=\frac{1}{2}(n^\mu-\bar n^\mu).
\end{align}
We have already defined two null vectors $n^\mu$ and $\bar n^\mu$ 
\bea 
n^\mu=(1,n^i),\quad \bar n^\mu=(-1,n^i) \label{null}
\eea with $n^i$ the normal vector of the unit sphere 
\be 
n^i=(\sin\theta\cos\phi,\sin\theta\sin\phi,\cos\theta).
\ee 
We further introduce $m^\m=\frac{1}{2}(n^\mu+\bar n^\mu)=(0,n^i)$ and $Y^A_\m=-\nabla^An_\mu$ where the covariant derivative $\nabla^A$ is adapted to the metric on the null boundary
\begin{align}
    ds^2_{\ci^+}=ds^2_{S^2}=d\th^2+\sin^2\th d\phi^2\equiv \g_{AB}d\th^A d\th^B.
\end{align}
The 6 conformal Killing vectors on the sphere are collected as $Y^A_\mn=Y^A_\m n_\n-Y^A_\n n_\m$ whose properties can be found in the appendix of \cite{Liu:2023gwa}. With the above definitions, we can express the partial derivative in Cartesian coordinates as
\begin{align}
    \p_\m=-n_\m\p_u+m_\m\p_r-r^{-1}Y^A_\m\p_A
\end{align}
which is useful to compute the boundary equations of motion and fluxes in the following.

\subsection{Equation of motion}\label{eom}
The equation of motion of the Dirac field is 
\be 
\slashed\partial \Psi=0
\ee which can be solved in momentum space. We will solve this equation near $\mathcal{I}^+$ with the fall-off condition
\be 
\Psi=\frac{1}{r}\Psi^{(1)}+\frac{1}{r^2}\Psi^{(2)}+\mathcal{O}(r^{-3}).
\ee 
where 
\bea 
\Psi^{(1)}=\left(\begin{array}{c}\psi_a(u,\Omega)\\\varphi^{\dagger \dot a}(u,\Omega)\end{array}\right),\quad \Psi^{(2)}=\left(\begin{array}{c}\psi^{(2)}_a(u,\Omega)\\\varphi^{\dagger(2)\dot a}(u,\Omega)\end{array}\right).
\eea 
The coefficients $\psi_a,\varphi^{\dagger \dot a},\psi_a^{(2)},\varphi^{\dagger(2)\dot a}$ are boundary fields that depend on the retarded coordinates of $\mathcal{I}^+$. 

We may switch the null vectors \eqref{null} to the spinor helicity formalism by defining
\be 
n_{a\dot a}=n_\mu\sigma^\mu_{a\dot a}=-\lambda_a\lambda^\dagger_{\dot a},\quad \bar{n}_{a\dot a}=\bar n_\mu\sigma^\mu_{a\dot a}=\kappa_a\kappa^\dagger_{\dot a}
\ee where the twistors $\lambda_a,\kappa_a$ and $\lambda^\dagger_{\dot a},\kappa^\dagger_{\dot a}$ may be chosen as
\bs\begin{align}
    \lambda_a&=i\sqrt{2}(\sin\frac{\theta}{2}e^{-i\phi/2},-\cos\frac{\theta}{2}e^{i\phi/2}),\quad \lambda^\dagger_{\dot a}=-i\sqrt{2}(\sin\frac{\theta}{2}e^{i\phi/2},-\cos\frac{\theta}{2}e^{-i\phi/2}),\\
    \kappa_a&=i\sqrt{2}(\cos\frac{\theta}{2}e^{-i\phi/2},\sin\frac{\theta}{2}e^{i\phi/2}),\quad \kappa_{\dot a}^\dagger=-i\sqrt{2}(\cos\frac{\theta}{2}e^{i\phi/2},\sin\frac{\theta}{2}e^{-i\phi/2}).
\end{align}\es 
One can find more properties on the twistors in Appendix \ref{twistor}.
With the commuting twistors, we can find 
\be 
\slashed n=\left(\begin{array}{cc}0&n_{a\dot a}\\ n^{\dot aa}&0\end{array}\right),
\ee where 
\bea 
n^{\dot aa}=\epsilon^{\dot a\dot b}\epsilon^{ab}n_{a\dot a}=n_\mu\bar{\sigma}^{\mu\dot a a}=-\lambda^{\dagger \dot a}\lambda^a.
\eea 
Then the equation of motion at order $\mathcal{O}(r^{-1})$ is 
\be 
n_{a\dot a}\varphi^{\dagger \dot a}=0,\quad n^{\dot a a}\psi_a=0.
\ee 
Note that the twistors $\lambda_a$ and $\kappa_a$ are linearly independent, we can always expand the two-component spinors $\psi_a,\varphi^{\dagger \dot a}$ as 
\be 
\psi_a=\lambda_a F+\kappa_a \tilde{F},\quad \varphi^{\dagger \dot a}=\lambda^{\dagger \dot a}\bar{G}+\kappa^{\dagger \dot a}\tilde{\bar{G}}.
\ee Using the identities in \eqref{contrac}, we can strip off the terms associated with $\kappa_a$ and $\kappa^{\dagger\dot a}$
\bea 
\psi_a=\lambda_a F,\quad \varphi^{\dagger\dot a}=\lambda^{\dagger\dot a}\bar G.
\eea Note that $F$ and $\bar G$ are independent fields on $\mathcal{I}^+$ which encode the radiative modes of the Dirac field. The subleading equation of motion is 
\bea 
\gamma^\mu[n_\mu\frac{\partial}{\partial u}\left(\begin{array}{c}\psi^{(2)}_a(u,\Omega)\\\varphi^{\dagger(2)\dot a}(u,\Omega)\end{array}\right)+m_\mu\left(\begin{array}{c}\psi_a(u,\Omega)\\\varphi^{\dagger \dot a}(u,\Omega)\end{array}\right)+Y^A_\mu\nabla_A\left(\begin{array}{c}\psi_a(u,\Omega)\\\varphi^{\dagger \dot a}(u,\Omega)\end{array}\right)]=0.
\eea 
We can expand the second order fields 
\be 
\psi_a^{(2)}=\lambda_a F^{(2)}+\kappa_a \tilde{F}^{(2)},\quad \varphi^{\dagger(2)\dot a}=\lambda^{\dagger \dot a}\bar G^{(2)}+\kappa^{\dagger \dot a}\tilde{\bar G}^{(2)}
\ee 
and utilize the identities \eqref{comb} to obtain 
\bs\begin{align}
   \dot{ \tilde{F}}^{(2)}&=-\frac{1}{2}(\zeta^A\nabla_AF+\frac{1}{2}\nabla_A\zeta^A F),\\
   \dot{ \tilde{\bar G}}^{(2)}&=-\frac{1}{2}(\bar\zeta^A\nabla_A\bar G+\frac{1}{2}\nabla_A\bar\zeta^A\bar G).
\end{align}\es 

\subsection{Canonical quantization}\label{cqz}
In the method of canonical quantization, the Dirac field can be written as the mode expansion \cite{2007qft..book.....S}
\bea 
\Psi(x)=\sum_{s=\pm}\int \frac{d^3\bm p}{(2\pi)^3 2\omega}[b_{s,\bm p} u_s(\bm p)e^{ip\cdot x}+d_{s,\bm p}^\dagger v_{s}(\bm p)e^{-ip\cdot x}]
\eea where $u_s(\bm p)$ and $v_s(\bm p)$ satisfy the equation of motion 
\be 
\slashed p u_s(\bm p)=\slashed p v_s(\bm p)=0.\label{uveom}
\ee The four-momentum $p_\mu$ is null 
\be 
p_\mu=\omega n_\mu,
\ee and we can switch it to the matrix form 
\be 
p_{a\dot a}=p_\mu \sigma^\mu_{a\dot a}=\omega n_{a\dot a}=-\omega \lambda_a\lambda^\dagger_{\dot a}.
\ee Therefore, the solution \eqref{uveom} becomes
\bea 
u_+(\bm p)=v_-(\bm p)=\sqrt{\omega}\left(\begin{array}{c}0\\ \lambda^{\dagger\dot a}\end{array}\right),\quad u_-(\bm p)=v_+(\bm p)=\sqrt{\omega}\left(\begin{array}{c} \lambda_{a}\\ 0\end{array}\right).
\eea The annihilation and creation operators $b_{s,\bm p}, d^\dagger_{s,\bm p}$ satisfy the anticommutators 
\bea 
\{b_{s,\bm p},b^\dagger_{s',\bm p'}\}=\{d_{s,\bm p},d^\dagger_{s',\bm p'}\}=(2\pi)^3 2\omega \delta_{s,s'}\delta(\bm p-\bm p'),
\eea while the vanishing anticommutators are omitted.
We expand the plane wave as the superposition of the spherical waves and find the mode expansion of the boundary fields 
\bs\label{modeFG}\begin{align}
    F(u,\Omega)&=-\frac{i}{8\pi^2}\int_0^\infty d\omega \sqrt{\omega}(b_{-,\bm p}e^{-i\omega u}-d_{+,\bm p}^\dagger e^{i\omega u}),\\
    \bar{G}(u,\Omega)&=-\frac{i}{8\pi^2}\int_0^\infty d\omega \sqrt{\omega}(b_{+,\bm p}e^{-i\omega u}-d_{-,\bm p}^\dagger e^{i\omega u}).
\end{align}\es We find the non-vanishing anticommutators 
\bs\label{anti}\begin{align}
    \{F(u,\Omega),\bar{F}(u',\Omega')\}&=\frac{1}{2}\delta(u-u')\delta(\Omega-\Omega'),\\
    \{G(u,\Omega),\bar{G}(u',\Omega')\}&=\frac{1}{2}\delta(u-u')\delta(\Omega-\Omega').
\end{align}\es 
There is no cross-talk term 
\be 
\alpha(u-u')=\frac{1}{2}[\theta(u'-u)-\theta(u-u')]
\ee in the anticommutators which contrasts with the commutators for bosonic fields \cite{Liu:2022mne}. It follows that there should be no non-local terms for the commutators between quantum flux operators. This fact is crucial for later discussions. At this moment, we can define the free vacuum state $|0\rangle$ as 
\be 
b_{s,\bm p}|0\rangle=d_{s,\bm p}|0\rangle=0
\ee and obtain the following correlators
\bs\begin{align}
    \langle 0|F(u,\Omega)\bar F(u',\Omega')|0\rangle&=-\frac{i}{4\pi(u-u'-i\epsilon)}\delta(\Omega-\Omega'),\\
    \langle 0|G(u,\Omega)\bar G(u',\Omega')|0\rangle&=-\frac{i}{4\pi(u-u'-i\epsilon)}\delta(\Omega-\Omega').
\end{align}\es 

\subsection{Quantum flux operators}\label{qfo}
In this subsection, we will use the stress tensor 
\be 
T^{\mu\nu}=-\frac{i}{4}(\bar\Psi \gamma^\mu\partial^\nu\Psi+\bar\Psi \gamma^\nu\partial^\mu\Psi-\partial^\nu\bar\Psi\gamma^\mu\Psi-\partial^\mu\bar\Psi\gamma^\nu\Psi)
\ee  to obtain the energy-momentum and angular momentum fluxes across $\mathcal{I}^+$. We choose a constant $r$ slice $\mathcal{H}_r$ and then compute the flux associated with the Killing vector $\bm \xi$ in the limit $\mathcal{H}_r\to \mathcal{I}^+$
\bea 
\mathcal{F}_{\bm\xi}=\lim_{r\to\infty,\ u\ \text{finite}}\int_{\mathcal{H}_r}(d^3x)_\mu T^{\mu}_{\ \nu}\xi^\nu.
\eea It turns out that the energy (and momentum) flux and the angular momentum (and center-of-mass) flux  are determined by the following flux density operators at  $\mathcal{I}^+$
\bs\begin{align}
T(u,\Omega)&=i(:\bar F\dot F-\dot{\bar F}F+G\dot{\bar G}-\dot G\bar{G}:),\\
S_A(u,\Omega)&=i(:\bar F\nabla_AF-\nabla_A\bar F F+G \nabla_A\bar G-\nabla_A G \bar G:),\\
O(u,\Omega)&=2(:\bar F F-G \bar G:).
\end{align}
\es 
Note that the left-handed and right-handed modes decouple with each other in the expression. Thus we can only consider the left-handed modes without loss of generality. We can construct three smeared operators as follows 
\bs\begin{align}
    \mathcal{T}_f&=\int du d\Omega f(u,\Omega)T(u,\Omega),\\
    \mathcal{M}_{\bm Y}&=\int du d\Omega [ Y^A(\Omega)S_A(u,\Omega)-\frac{i}{8}o(y,\bar y)O(u,\Omega)],\\
    \mathcal{O}_{h}&=\int du d\Omega h(u,\Omega)O(u,\Omega)
\end{align}\es where $f(u,\Omega)$ and $h(u,\Omega)$ are arbitrary smooth functions on $\mathcal{I}^+$ and $Y^A(\Omega)$ is any smooth vector field on $S^2$. The vector $Y^A$ can be decomposed into two modes $y,\bar y$ using the vectors $\zeta^A,\bar\zeta^A$ in \eqref{zeta}
\be 
Y^A=\frac{1}{2}(y\bar\zeta^A+\bar y\zeta^A),\quad y=Y^A\zeta_A,\quad \bar y=Y^A\bar\zeta_A.
\ee The function $o(y,\bar y)$ is 
\be 
o(y,\bar y)=\zeta^A\nabla_A \bar y-\bar\zeta^A\nabla_Ay+\nabla_A\bar\zeta^A y-\nabla_A\zeta^A \bar y.
\ee
The physical meaning of the above three flux operators are shown in Table \ref{phy}.
\begin{table}
\begin{center}
\renewcommand\arraystretch{1.5}
    \begin{tabular}{|c||c|c|c|}\hline
Functions $(f,\bm Y,h)$&$\mathcal{T}_f$&$\mathcal{M}_{\bm Y}+\mathcal{T}_f$&$\mathcal{O}_h$\\\hline\hline
$(1,0,0)$&\text{energy flux}&&\\\hline
$(n^i,0,0)$&\text{momentum flux}&&\\\hline
$(0,-Y^A_{ij},0)$&&\text{angular momentum flux}&\\\hline
$(-un^i,-Y^A_i,0)$&&\text{center-of-mass flux}&\\\hline
$(0,0,1)$&&&\text{helicity flux}\\\hline
\end{tabular}
\caption{\centering{Correspondence between the test functions and the fluxes. }}\label{phy}
\end{center}
\end{table}
As in the bosonic case, $\mathcal{T}_f$ and $\mathcal{M}_{\bm Y}$ may be regarded as energy and angular momentum fluxes, respectively. 

\paragraph{Helicity flux operator and chiral symmetry.}
The helicity flux operator $\mathcal{O}_{h=1}$ is associated with the chiral symmetry for the massless Dirac theory 
\bea 
\Psi\to \Psi'=e^{i\theta_5\gamma_5}\Psi,\quad \bar\Psi\to \bar\Psi'=\bar\Psi e^{i\theta_5\gamma_5}
\eea where $\theta_5$ is a constant and $\gamma_5$ is 
\be 
\gamma_5=i\gamma^0\gamma^1\gamma^2\gamma^3=\left(\begin{array}{cc}-1_{2\times 2}&0\\0&1_{2\times 2}\end{array}\right)
\ee which is anticommuting with $\gamma^\mu$ matrices
\be 
\gamma^5\gamma^\mu=-\gamma^\mu\gamma^5.
\ee The axial current of the chiral symmetry is 
\be 
j_5^\mu=\bar\Psi\gamma^\mu\gamma_5\Psi.
\ee Classically,  the axial current is 
conserved for free massless Dirac theory. Now we will evaluate the same quantity using the axial current
\bea 
\mathcal{F}_{\text{chiral}}&=&\lim_{r\to\infty,\ u\ \text{finite}}\int_{\mathcal{H}_r} (d^3x)_\mu j^\mu_5\nn\\&=&-\int du d\Omega m_\mu[\varphi\sigma^\mu\varphi^\dagger-\psi^\dagger \bar\sigma^\mu\psi]\nn\\&=&2\int du d\Omega (\bar FF-G\bar G).
\eea
This is the helicity flux operator with $h=1$.
Recall the mode expansion \eqref{modeFG}, the helicity flux can be written as
\be \mathcal{F}_{\text{chiral}}=\int \frac{d^3\bm p}{(2\pi)^3 2\omega}(b_{-,\bm p}^\dagger b_{-,\bm p}+d_{-,\bm p}^\dagger d_{-,\bm p}-b_{+,\bm p}^\dagger b_{+,\bm p}-d_{+,\bm p}^\dagger d_{+,\bm p})\label{fchiral}
\ee which is exactly the difference of the Dirac particle numbers with negative and positive helicities. Note that the axial current is not conserved for QED as well as other gauge theories with massless spinors at the quantum level due to the chiral anomaly \cite{Adler:1969gk,Bell:1969ts}. However, the definition of the helicity flux operator is independent of the chiral anomaly.  We will discuss the relation between the anomaly and the helicity flux operator in the conclusion.

\subsection{Supertranslation and superrotation generators}\label{stsr}
In this subsection, we will show that the quantum flux operators $\mathcal{T}_f$ and $\mathcal{M}_{\bm Y}$ can also be interpreted as supertranslation and superrotation generators respectively for $f$ and $\bm Y$ are time-independent. To prove this point, we should derive the variation of the fields $F, G$ under supertranslation and superrotation.

\paragraph{Commutators between the quantum flux operators and the boundary fields.}
Using the anticommutators \eqref{anti}, we find
\bs\begin{align}
    [\mathcal{T}_f,F(u,\Omega)]&=-i f(u)\dot F(u)-\frac{i}{2}\dot f(u)F(u),\label{stF}\\
    [\mathcal{T}_f,\bar F(u,\Omega)]&=-i f(u)\dot {\bar{F}}(u)-\frac{i}{2}\dot f(u)\bar{F}(u),\label{stbF}\\
    [\mathcal{M}_{\bm Y},F(u,\Omega)]&=-iY^A(u)\nabla_AF(u)-\frac{i}{2}\nabla_AY^A(u) F(u)+\frac{i}{8}o(y,\bar y)F(u)\label{srF},\\
     [\mathcal{M}_{\bm Y},\bar F(u,\Omega)]&=-iY^A(u)\nabla_A\bar{F}(u)-\frac{i}{2}\nabla_AY^A(u) \bar{F}(u)-\frac{i}{8}o(y,\bar y)\bar F(u),\\
    [\mathcal{O}_h,F(u,\Omega)]&=-h(u)F(u),\\
     [\mathcal{O}_h,\bar F(u,\Omega)]&=h(u)\bar{F}(u).
\end{align}\es  The transformation \eqref{stF} indicates that $F(u,\Omega)$ is the primary operator with weight $1/2$, which is different from boundary bosonic fields. Note that the commutators are valid for any time-dependent test functions $f,\bm Y,h$. On the right hand side, we have used the abbreviation
\be 
f(u,\Omega)=f(u),\quad Y^A(u,\Omega)=Y^A(u),\quad h(u,\Omega)=h(u).
\ee Unlike the bosonic field,  there are no non-local terms. The variations should match with the variations induced by bulk Lie derivative of the spinor field in certain limit. 
\paragraph{Lie derivative of a spinor field.} Interestingly, the Lie derivative of a spinor field along a given vector field $\bm\xi$ is not uniquely defined in general. In 1963, Lichnerowicz gave the first definition of the Lie derivative of a spinor field along a Killing vector. This work was generalized by Kosmann  \cite{Kosmann1971DrivesDL} which is now the  most commonly used Lie derivative of a spinor field.  Penrose and Rindler gave another extension which is suitable for the Lie derivative of the spinor field along any  conformal Killing vectors \cite{Penrose:1986ca}. One can find more discussion on this topic in \cite{Habermann1996TheGA,Godina:2005mt,Leao:2015qqc}.
 To determine which definition is suitable for our problem, we may introduce a one-parameter family of the 
 Lie derivative of a spinor field along $\bm \xi$\footnote{ More discussions on the ambiguity on the Lie derivative of the spinor field can be found in \cite{Godina:2003tc}. The one-parameter family of the Lie derivative of the spinor field has been explicitly written out in  \cite{Leao:2015qqc,Helfer:2016zvl}.}
\be 
\mathcal{L}_{\bm\xi}\Psi=\xi^\mu\nabla_\mu\Psi-\frac{1}{4}\nabla_{[\mu}\xi_{\nu]}\gamma^\mu\gamma^\nu\Psi+\alpha\nabla_\mu\xi^\mu \Psi,\label{Liederivative}
\ee where the covariant derivative of the spinor field $\Psi$ is 
\be 
\nabla_\mu\Psi=\partial_\mu\Psi-\frac{1}{4}\omega_\mu^{\ ab}\gamma_a\gamma_b\Psi
\ee 
with $\gamma^a$ the Dirac gamma matrices in the Cartesian frame 
\be 
\gamma^a=e^a_\mu\gamma^\mu.
\ee The vielbein fields $e^a_\mu$ obey the orthogonality and completeness relations 
\be 
e_\mu^a e_\nu^b g^{\mu\nu}=\eta^{ab},\quad e_\mu^a e_\nu^b \eta_{ab}=g_{\mu\nu},
\ee which can be used to determine the spin connection $\omega_{\mu}{}^{ab}$ through the torsion free condition 
\bea 
\partial_\mu e_\nu^a-\partial_\nu e_\mu^a+\omega_{\mu a}{}^b e_{\nu b}-\omega_{\nu a}{}^b e_{\mu b}=0.
\eea 
We find that $\alpha=0$ corresponds to Kosmann's original definition of Lie derivative. 

Taking the derivatives $x^\mu=u \bar m^\mu+r n^\mu$, we find the following relations 
\be 
du=-n_\mu dx^\mu,\quad dr=m_\mu dx^\mu,\quad d\theta^A=-\frac{1}{r}Y_\mu^A dx^\mu.
\ee Then the metric \eqref{metric} becomes 
\bea 
ds^2=\eta_{\mu\nu}dx^\mu dx^\nu.
\eea 

Therefore, we can choose the vielbein as 
\be 
e_\mu^a=\delta_\mu^a\quad\Rightarrow\quad e^a=e_\mu^adx^\mu=\delta_\mu^a(\bar m^\mu du+n^\mu dr-r Y^\mu_A d\theta^A)
\ee such that the spin connection
 is always zero. 

\paragraph{Supertranslation generator.}
Transformations generated by the vector 
\bea 
\bm \xi_f&=&f\partial_u+\frac{1}{2}\nabla_A\nabla^A f\partial_r-\frac{1}{r}\nabla^A f\partial_A+\mathcal{O}(r^{-1})\nn\\
&=&[f\bar m^\mu+\frac{1}{2}\nabla^2fn^\mu+Y^\mu_A \nabla^A f+\mathcal{O}(r^{-1})]\partial_\mu\label{GSTs}
\eea are called supertranslations. 
The function $f=f(\Omega)$ is  smooth on $S^2$. Utilizing the Lie derivative \eqref{Liederivative}, we find the variation of the boundary fields $F,\bar G$ under supertranslation
\be
\delta_f F(u,\Omega)=f(\Omega)\dot{F}(u,\Omega),\quad \delta_f \bar{G}(u,\Omega)=f(\Omega)\dot{\bar G}(u,\Omega).
\ee On the other hand, we can easily find  that \eqref{stF} is reduced to 
\be 
[\mathcal{T}_f,F(u,\Omega)]=-if(\Omega)\dot{F}(u,\Omega)
\ee once $f$ is independent of time. We conclude that $\mathcal{T}_f$ is the generator of supertranslation for $f=f(\Omega)$. {Note that we do not need to specify the choice of $\a$ at this moment.}

Now we turn to the case that $f$ is dependent on $u$. Using Lie derivative \eqref{Liederivative}, we would obtain 
\bs\begin{align} 
\delta_fF(u,\Omega)&=f(u,\Omega)\dot F(u,\Omega)+(\frac{1}{4}+\alpha)\dot{f}(u,\Omega)F(u,\Omega),\\
\delta_f \bar G(u,\Omega)&=f(u,\Omega)\dot {\bar G}(u,\Omega)+(\frac{1}{4}+\alpha)\dot{f}(u,\Omega)\bar G(u,\Omega).
\end{align}\es
In this case, the variation of $F$ does not match with the commutator \eqref{stF} for general $\alpha$ unless
\be 
\alpha=\frac{1}{4}.
\ee  
{This is different from the choices of Kosmann or Penrose and Rindler.}

\paragraph{Superrotation generator.}
Now we consider the 
superrotations that is generated by the vector 
\begin{align}
    \bm \xi_{\bm Y}&=\frac{1}{2}u\nabla_AY^A\partial_u-\frac{1}{2}r\nabla_AY^A\partial_r+\frac{u }{4}\nabla_C \nabla^C\nabla\cdot Y\partial_r+(Y^A-\frac{u}{2r}\nabla^A\nabla\cdot Y)\partial_A+\mathcal{O}(r^{-1})\nn\\
    &=\left(-rY^AY^\mu_A-\frac{1}{2}r \nabla\cdot Y n^\mu+\frac{1}{2}u \nabla\cdot Y \bar m^\mu+\frac{1}{4}u \nabla^2\nabla\cdot Y n^\mu+\frac{1}{2}u \nabla^A \nabla\cdot Y Y^\mu_A+\mathcal{O}(r^{-1})\right)\partial_\mu,
\end{align}
where $Y^A(\Omega)$ is any smooth vector on $S^2$. Now we can find the variation of the boundary fields under superrotation 
\bs\begin{align}
    \delta_{\bm Y}F(u,\Omega)&=\frac{1}{2}u\nabla\cdot Y \dot F+\frac{1}{4}\nabla\cdot Y F+Y^A\nabla_A F+\frac{1}{2}\nabla\cdot Y F-\frac{1}{8}o(y,\bar y)F,\label{deltayF}\\
    \delta_{\bm Y}\bar G(u,\Omega)&=\frac{1}{2}u \nabla\cdot Y \dot{\bar G}+\frac{1}{4}\nabla\cdot Y \bar G+Y^A\nabla_A\bar G+\frac{1}{2}\nabla\cdot Y\bar G+\frac{1}{8}o(y,\bar y)\bar G.
\end{align}\es
Note that the variation is independent of $\alpha$ since $\nabla_\mu\xi_{\bm Y}^\mu$ falls off as $r^{-1}$. Note that the first two terms are exactly the variation 
\be 
\delta_{f=\frac{1}{2}u\nabla\cdot Y}F
\ee for $\alpha=\frac{1}{4}$. The last three terms match with the commutator \eqref{srF}. Therefore, we find the identity 
 \bea  
 \delta_{\bm Y}F(u,\Omega)-\delta_{f=\frac{1}{2}u\nabla\cdot Y}F(u,\Omega)= [i\mathcal{M}_{\bm Y},F(u,\Omega)]\label{2.79}
\eea which is exactly the same as the bosonic theories \cite{Liu:2024nkc}.
For all other values of $\alpha$, the variations induced from bulk do not match with the boundary commutators, both for supertranslation and superrotation. 

\paragraph{Other variations.}
The variations of the fields $\bar F$ and $G$ may be obtained from the complex conjugation. However, one may also use the Lie derivative of the spinor field $\bar \Psi$ 
\bea 
\mathcal{L}_{\bm\xi}\bar\Psi=\xi^\mu\nabla_\mu\bar\Psi+\frac{1}{4}\nabla_{[\mu}\xi_{\nu]}\bar\Psi\gamma^\mu\gamma^\nu+\alpha\nabla_\mu\xi^\mu\bar\Psi.\label{lie}
\eea 
Note that to obtain above expression, we should take the Hermitian conjugate and assume that the Lie derivatives of the invariant form $\epsilon_{ab}$ and $\epsilon_{\dot a\dot b}$ are zero
\bea 
\mathcal{L}_{\bm\xi}\epsilon_{ab}=\mathcal{L}_{\bm\xi}\epsilon_{\dot a\dot b}=\mathcal{L}_{\bm\xi}\epsilon^{ab}=\mathcal{L}_{\bm\xi}\epsilon^{\dot a\dot b}=0.\label{invariantform}
\eea The above condition is motivated by the identity 
\bea 
\epsilon_{ab}=\frac{1}{2}(\lambda_a\kappa_b-\kappa_a\lambda_b),\quad \epsilon_{\dot a\dot b}=\frac{1}{2}(\lambda^\dagger_{\dot a}\kappa^\dagger_{\dot b}-\lambda^\dagger_{\dot b}\kappa^\dagger_{\dot a}).
\eea Since the twistors are fixed, they are invariant under Lie derivatives \be 
\lambda'_a(\Omega')=\lambda(\Omega'),\quad \kappa_a'(\Omega')=\kappa_a(\Omega'),\quad \lambda'^\dagger_{\dot a}(\Omega')=\lambda^\dagger(\Omega'),\quad \kappa'^{\dagger}_{\dot a}(\Omega')=\kappa^\dagger_{\dot a}(\Omega'),
\ee which lead to \eqref{invariantform}. These conditions could compare with the assumption that the covariant variation of the boundary metric $\gamma_{AB}$ vanishes for bosonic theory \cite{Liu:2023qtr} 
\be 
\slashed\delta_{\bm Y}\gamma_{AB}=0.\label{covgamma}
\ee 
The boundary metric $\gamma$ is used to raise and lower the indices of the boundary field $A_A$. For any scattering process, we should fix this boundary metric\footnote{The boundary metric can be constructed by the vielbein field $e_A^a$ on the celestial sphere which is related to  the polarization vector \cite{Liu:2024llk}. Normally, we will fix the form of the polarization vector and therefore fix the boundary metric. }, and this leads to the condition \eqref{covgamma}. This is exactly similar to the fermion case that we use $\epsilon$ to raise and lower spinor indices. 

We should mention that there could be another Lie derivative for the field $\bar\Psi$ 
\bea 
\tilde{\mathcal{L}}_{\bm\xi}\bar\Psi=\xi^\mu\nabla_\mu\bar\Psi+\frac{1}{4}\nabla_{[\mu}\xi_{\nu]}\bar\Psi\gamma^\mu\gamma^\nu-\alpha\nabla_\mu\xi\bar\Psi,
\eea which is to flip the sign in the last term. This definition is to require that $\bar\Psi\Psi$ is a scalar 
\be 
\tilde{\mathcal{L}}_{\bm\xi}(\bar\Psi\Psi)=\xi^\mu\nabla_\mu(\bar\Psi\Psi)
\ee and assume the Leibniz rule for the Lie derivative. As a consequence, the invariant form $\epsilon$ will transform as 
\bs\begin{align}
    \tilde{\mathcal{L}}_{\bm\xi}\epsilon^{ab}&=-2\alpha\nabla\cdot\xi \epsilon^{ab},\quad \tilde{\mathcal{L}}_{\bm\xi}\epsilon_{ab}=2\alpha\nabla\cdot\xi\epsilon_{ab},\\
    \tilde{\mathcal{L}}_{\bm\xi}\epsilon^{\dot a\dot b}&=2\alpha\nabla\cdot\xi \epsilon^{\dot a\dot b},\quad \tilde{\mathcal{L}}_{\bm\xi}\epsilon_{\dot a\dot b}=-2\alpha\nabla\cdot\xi \epsilon_{\dot a\dot b}.
\end{align}\es 
Therefore, this definition is not consistent with the invariance of the twistors. Moreover, we would obtain 
\be 
\tilde\delta_{f}\bar F=f(u,\Omega)\dot{\bar F}(u,\Omega)
\ee 
{for $\alpha=1/4$ which} is also inconsistent with the commutator \eqref{stbF}. 

\subsection{Flux algebra and subalgebras}\label{icd}
Since there is no non-local term in the commutator between the boundary fields $F$ and $\bar F$, it would be interesting to 
compute the intertwined Carrollian  diffeomorphism where all the test functions $f, Y^A,h$ are time and angle dependent. After taking into account the contribution from $G$ and $\bar G$, the result is as follows
\bs\label{spinoralgebra}
\begin{align}
[\mathcal{T}_{f_1},\mathcal{T}_{f_2}]&=C_T(f_1,f_2)+i\mathcal{T}_{f_1\dot{f}_2-f_2\dot{f_1}},\label{Tf1f2}\\
[\mathcal{T}_{f},\mathcal{M}_{\bm Y}]&=C_{TM}(f, \bm Y)-i\mathcal{T}_{\bm Y(f)}+i\mathcal{M}_{f\dot{Y}^A}- \frac{i}{4}\mathcal{O}_{\epsilon_{BA}\dot{Y}^A\nabla^B f}  ,\label{TfMY}\\
[\mathcal{T}_f,\mathcal{O}_h]&=i\mathcal{O}_{f\dot h},\\
[\mathcal{M}_{\bm Y},\mathcal{M}_{\bm Z}]&=C_M(\bm Y,\bm Z)+i\mathcal{M}_{[\bm Y,\bm Z]}-\frac{i}{2}\mathcal{O}_{o(\bm Y,\bm Z)},\label{MYZ}\\
[\mathcal{M}_{\bm Y},\mathcal{O}_{h}]&=C_{MO}(\bm Y,h)+i\mathcal{O}_{\bm Y(h)},\\
[\mathcal{O}_{h_1},\mathcal{O}_{h_2}]&=C_O(h_1,h_2),
\end{align}\es 
where we have used the identities
\bs\begin{align}
&o(f\dot{y},f\dot{\bar y})-fo(\dot y,\dot{\bar y})=(\dot{\bar y}\zeta^A-\dot{y}\bar{\zeta}^A)\nabla_Af=-2i \epsilon_{AB}\dot{Y}^A\nabla^Bf,\\
&Y^A\nabla_A o(z,\bar z)-Z^A\nabla_Ao(y,\bar y)=o([\bm Y,\bm Z]^A\zeta_A,[\bm Y,\bm Z]^A\bar \zeta_A)-4io(\bm Y,\bm Z).
\end{align}\es 
The notations in the algebra are defined as follows
\bs\begin{align}
&\bm Y(h)=Y^A\nabla_A h,\qquad [\bm Y,\bm Z]^A=Y^B\nabla_BZ^A-Z^B\nabla_BY^A,\\
&o(\bm Y,\bm Z)=\frac{1}{4}\epsilon^{BC}\Theta_{AB}(\bm Y)\Theta^A_{\ C}(\bm Z),\quad \Theta_{AB}(\bm Y)=\nabla_{A}Y_B+\nabla_BY_A-\gamma_{AB}\nabla_CY^C.
\end{align}\es 
The algebra \eqref{spinoralgebra} is akin to the bosonic cases, except for some novel points. 
\paragraph{Central charges}
In the commutators, the central charges are 
\bs\begin{align}
    C_T(f_1,f_2)&=-\frac{ic}{24\pi}\mathcal{I}_{f_1\dddot{f}_2-f_2\dddot{f}_1},\\
    C_{M}(\bm Y,\bm Z)&=\frac{i}{2\pi}\int du d\Omega d\Omega'\Lambda_{AB'}(\Omega,\Omega')Y^A(u,\Omega)\dot {Z}^{B'}(u,\Omega')-\frac{1}{64}C_O(o(y,\bar y),o(z,\bar z)),\\
    C_{MO}(\bm Y,h)&=\frac{c}{16\pi}\mathcal{I}_{o(y,\bar y)\dot h-o(\dot y,\dot{\bar y})h},\\
    C_O(h_1,h_2)&=\frac{ic}{2\pi}\mathcal{I}_{h_1\dot{h}_2-h_2\dot{h}_1},
\end{align}\es where $\mathcal I_f$ is defined as
\bea 
\mathcal I_f=\int du d\Omega f(u,\Omega).
\eea and $\Lambda_{AB'}(\Omega,\Omega')$ is 
\be 
\Lambda_{AB'}(\Omega,\Omega')=\delta(\Omega-\Omega')\nabla_A\nabla_{B'}\delta(\Omega-\Omega')-\nabla_A\delta(\Omega-\Omega')\nabla_{B'}\delta(\Omega-\Omega').
\ee The constant $c=\delta^{(2)}(0)$ is the divergent Dirac delta function on the unit sphere which may be regularized \cite{Li:2023xrr} using zeta function or heat kernel method.
The structure $\Lambda_{AB'}(\Omega,\Omega')$ have already appeared in the scalar theory. As expected, there are no annoying non-local terms in the fermionic theory. 

\paragraph{Intertwined Carrollian diffeomorphism.}
We notice that the central charge of the superrotation generators is  distinguished from other three central charges. Recall that a generalized superrotation with time-dependent $\bm Y$ will break the null structure of the Carrollian manifold $\mathcal{I}^+$, we may rule out this possibility and set $\dot {\bm Y}=0$. Actually, there is another technical difficulty for the generalized function $\Lambda_{AB'}$ since it contains the product of the Dirac function and its derivatives with respect to the angles. Till now, we do not find a method to regularize this quantity. After imposing this condition, we obtain the following algebra
\bs\label{spinoralgebrap}
\begin{align}
[\mathcal{T}_{f_1},\mathcal{T}_{f_2}]&=C_T(f_1,f_2)+i\mathcal{T}_{f_1\dot{f}_2-f_2\dot{f_1}},\label{Tf1f2p}\\
[\mathcal{T}_{f},\mathcal{M}_{\bm Y}]&=-i\mathcal{T}_{\bm Y(f)},\label{TfMYp}\\
[\mathcal{T}_f,\mathcal{O}_h]&=i\mathcal{O}_{f\dot h},\\
[\mathcal{M}_{\bm Y},\mathcal{M}_{\bm Z}]&=i\mathcal{M}_{[\bm Y,\bm Z]}-\frac{i}{2}\mathcal{O}_{o(\bm Y,\bm Z)},\label{MYZp}\\
[\mathcal{M}_{\bm Y},\mathcal{O}_{h}]&=i\mathcal{O}_{\bm Y(h)},\\
[\mathcal{O}_{h_1},\mathcal{O}_{h_2}]&=C_O(h_1,h_2)\label{OOp}
\end{align}\es with 
\be 
 C_T(f_1,f_2)=-\frac{ic}{24\pi}\mathcal{I}_{f_1\dddot{f}_2-f_2\dddot{f}_1},\quad
    C_O(h_1,h_2)=\frac{ic}{2\pi}\mathcal{I}_{h_1\dot{h}_2-h_2\dot{h}_1}.\label{ctco}
\ee 
Note that time-dependent $h$  is self-consistent even though $\bm Y$ and $\bm Z$ are time-independent, in contrast with the bosonic theory where the right hand side of the commutator  \eqref{OOp} would have non-local operators. We will regard the above algebra \eqref{spinoralgebrap} the fermionic generalization of the intertwined Carrollian diffeomorphism. When $h$ is also time-independent, the algebra reduces to the standard Carrollian diffeomorphism \cite{Liu:2024nkc} with the spin $s=\frac{1}{2}$.\footnote{Note that we should flip the sign of $\mathcal{O}$ in the definition to match with the bosonic result.}

{
\paragraph{Jacobi identity.}
We find that the nontrivial central charges \eqref{ctco} will lead to the violation of Jacobi identities for the flux operator triples $\ct\ct\cm$ and  $\co\co\cm$
\begin{subequations}\label{viodirac}
  \begin{align}
  J[\ct_{f_1},\ct_{f_2},\cm_{\bm Y}]&=-\frac{c}{24\pi}\ci_{Y^A\nabla_A(f_1\p_u^3f_2-f_2\p_u^3f_1)},\\ 
  J[\mathcal{O}_{h_1},\mathcal{O}_{h_2},\mathcal{M}_{\bm Y}]&=\frac{c}{2\pi}\ci_{Y^A\nabla_A(h_1\dot h_2-h_2\dot h_1)},
\end{align}\label{jacvio}
\end{subequations}
where the Jacobiator is defined as
    \begin{align}
        J[x,y,z]=[x,[y,z]]+[y,[z,x]]+[z,[x,y]].
    \end{align}
    
    There are several ways to satisfy the Jacobi identity. 
    
    At first, we may require the central charge $c=0$. However, we are not aware of any way to regularize $c$ to zero in even dimensions \cite{Li:2023xrr}. When there are multiple fields with different types, the total central charge is the addition of the individual central charges associated with each field.  Therefore, the total central charge may still be vanishing by selecting the field content in the theory. We do not find a solution in this way and will not discuss this possibility in this work.
    
    Second, we can constrain to a divergence-free vector $Y^A$ which makes the violation terms in \eqref{viodirac} total derivatives on the sphere thus vanishing. It is known that the Killing vector on the sphere is divergence-free, i.e., $\nabla_AY^A_{ij}=0$, so we can constrain $Y^A$ to Killing vectors although this is a bit weird since the Lorentz boost is excluded (and of course, there will be no helicity flux appearing in the commutator $[\cm_{\bm Y},\cm_{\bm Z}]$). More generally, since the commutator of two divergence-free vectors on the sphere is still divergence-free which can be seen from the identity
    \begin{align}
\nabla_A[\bm Y, \bm Z]^A=  Y^A\nabla_A\nabla_B Z^B-Z^B\nabla_B\nabla_AY^A,\label{ynnza.34}
\end{align}
we can allow $Y^A$ to take any divergence-free smooth vector field which is the extension of spatial rotation. It generates the diffeomorphism with odd parity which is the second term in the following decomposition \cite{Flanagan:2015pxa}
\begin{align}
    Y^A = \nabla^A \mathcal K + \epsilon^{AB} \nabla_B \mathcal Y.
\end{align}
As an example, our Killing vector $Y_{ij}^A$ can be recast to 
\begin{align}
   { Y^A_{ij}=\epsilon^{AB}\nabla_B (\epsilon_{ijk}n^k)}
\end{align}
and hence is magnetic. We call this divergence-free $Y^A$ magnetic superrotation which will also give rise to the helicity flux. 
The algebra is the same as \eqref{spinoralgebrap} with the constraint
\be 
Y^A=\epsilon^{AB}\nabla_B\mathcal Y
\ee where $\mathcal Y$ is an arbitrary smooth function on $S^2$. Note that the choice $Y^A=\nabla^A\mathcal K$ does not lead to a closed algebra. 
    
The third way is to constrain the test functions in the supertranslation and helicity flux operator.  We should at least require $\dddot{f}=0$ and $\dot h=0$ which reduce the flux algebra to
\begin{subequations}\label{pu3f}
  \begin{align}
    &[\mathcal{T}_{f_1},\mathcal{T}_{f_2}]=i\mathcal{T}_{f_1\dot f_2-f_2\dot f_1},&&
    [\mathcal{T}_{f},\mathcal{M}_{\bm Y}]=-i\mathcal{T}_{\bm Y(f)},\\
    &[\mathcal{T}_f,\mathcal{O}_h]=0,&&
    [\mathcal{M}_{\bm Y},\mathcal{M}_{\bm Z}]=i\mathcal{M}_{[\bm Y, \bm Z]}-\frac{i}{2}\co_{o(\bm Y, \bm Z)},\\
    &[\mathcal{M}_{\bm Y},\mathcal{O}_h]=i\mathcal{O}_{\bm Y(h)},&&
    [\mathcal{O}_{h_1},\mathcal{O}_{h_2}]=0.
  \end{align}
\end{subequations}

There are three kinds of solutions which are listed as follows.
\begin{itemize}
    \item One choice is to demand $\dot h=0$ and $f$ quadratic at $u$ \footnote{This is similar to the truncation of the supertranslation parameter of the Newman-Unti group to a polynomial of $u$. One can find more details in \cite{Duval_2014b}.}
\be 
f(u,\Omega)=T(\Omega)+u W(\Omega)+u^2 V(\Omega).
\ee 
This choice leads to the algebra \eqref{pu3f} with
\begin{align}
    f_1\dot f_2-f_2\dot f_1=u^2(W_1V_2-W_2V_1)+2u(T_1V_2-T_2V_1)+T_1W_2-T_2W_1
\end{align}
which will not break $\p_u^3f=0$. To be more explicit, we can split our generalized supertranslation flux as
\begin{align}
  \ct_f=\cp_T^0+\cp^1_W+\cp^2_V,
\end{align}
where
\begin{subequations}
  \begin{align}
    \cp_T^0&=\int dud\Omega\,T(\Omega)T(u,\Omega),\\
    \cp_W^1&=\int dud\Omega\,uW(\Omega)T(u,\Omega),\\
    \cp_V^2&=\int dud\Omega\,u^2V(\Omega)T(u,\Omega).
  \end{align}
\end{subequations}
Then we have the following structure
\begin{subequations}
  \begin{align}
    &[\cp_{T_1}^0,\cp_{T_2}^0]=[\cp_{W_1}^1,\cp_{W_2}^1]=[\cp_{V_1}^2,\cp_{V_2}^2]=0,\\ 
    &[\cp_{T}^0,\cp_{W}^1]=i\cp^0_{TW},\quad [\cp_{T}^0,\cp_{V}^2]=i\cp^1_{2TV},\quad [\cp_{W}^1,\cp_{V}^2]=i\cp^2_{WV}.
  \end{align}
\end{subequations}
 
\item A more significant choice is to demand $\dot h=0$ and $f$ linear at $u$ namely \(f(u,\Omega)=T(\Omega)+uW(\Omega)\). These conditions leads to an intertwined Weyl BMS (BMSW) algebra realized for the Dirac theory. To avoid duplication, we only list
\begin{align}
    [\mathcal{T}_{f_1},\mathcal{T}_{f_2}]=i\mathcal{T}_{T_1W_2-T_2W_1},
\end{align}
which is equivalent to
\begin{align}
    &[\cp_{T_1}^0,\cp_{T_2}^0]=[\cp_{W_1}^1,\cp_{W_2}^1]=0,\qquad [\cp_{T}^0,\cp_{W}^1]=i\cp^0_{TW}.
\end{align}
The original BMSW algebra \cite{Freidel:2021fxf} comes from relaxing the boundary conditions to $\cl_{\xi}g_{AB}=\co(r^2)$ as well as $\cl_{\xi}g_{uu}=\co(1)$ and consists of ${\rm Diff}(S^2)\ltimes C^\infty_T(S^2)\ltimes C^\infty_W(S^2)$. Note that our case is slightly different from the original BMSW algebra since we have a helicity flux operator in the algebra. However, the BMSW algebra exactly has the same structure as ours if excluding the helicity flux. 
\item We can further require $\dot f=0$ and $\dot h=0$ which gives an intertwined generalized BMS (gBMS) algebra
\begin{subequations}
    \begin{align}
    &[\mathcal{T}_{f_1},\mathcal{T}_{f_2}]=0,&&
    [\mathcal{T}_{f},\mathcal{M}_{\bm Y}]=-i\mathcal{T}_{\bm Y(f)},\\
    &[\mathcal{T}_f,\mathcal{O}_h]=0,&&
    [\mathcal{M}_{\bm Y},\mathcal{M}_{\bm Z}]=i\mathcal{M}_{[\bm Y, \bm Z]}-\frac{i}{2}\co_{o(\bm Y, \bm Z)},\\
    &[\mathcal{M}_{\bm Y},\mathcal{O}_h]=i\mathcal{O}_{\bm Y(h)},&&
    [\mathcal{O}_{h_1},\mathcal{O}_{h_2}]=0.
  \end{align}
\end{subequations}
\end{itemize}

Besides the above Lie algebra gained from constraining parameters, we can also find a closed Lie algebra by excluding $\cm_{\bm Y}$
  \begin{align}
    &[\mathcal{T}_{f_1},\mathcal{T}_{f_2}]=C_T(f_1,f_2)+i\mathcal{T}_{f_1\dot f_-f_2\dot f_1},\quad [\mathcal{T}_f,\mathcal{O}_h]=i\co_{f\dot h},\quad
    [\mathcal{O}_{h_1},\mathcal{O}_{h_2}]=C_O(h_1,h_2).
  \end{align} They form a higher-dimensional Kac-Moody algebra in Fourier space.\footnote{The construction of Kac-Moody algebra can be found in \cite{Moody1968ANC,1968IzMat...2.1271K} and their applications in physics can be found in the book on two-dimensional conformal field theories.} To see this point, we define the basis functions
  \be 
f_{\omega,\ell,m}=h_{\omega,\ell,m}=e^{-i\omega u}Y_{\ell,m}(\Omega).
\ee 
Then one can easily find
\bs\begin{align}
& [\mathcal{T}_{\omega,\ell,m},\mathcal{T}_{\omega',\ell',m'}]=(\omega'-\omega)\sum_{L,M} c_{\ell,m;\ell',m';L,M}\mathcal{T}_{\omega+\omega',L,M}-(-1)^m\frac{\omega^3}{6}c\ \delta(\omega+\omega')\delta_{\ell,\ell'}\delta_{m,-m'},\\
    &[\mathcal{T}_{\omega,\ell,m},\mathcal{O}_{\omega',\ell',m'}]=\omega'\sum_{L,M}c_{\ell,m;\ell',m';L,M}\mathcal{O}_{\omega+\omega',L,M},\\
    &[\mathcal{O}_{\omega,\ell,m},\mathcal{O}_{\omega',\ell',m'}]=-2c\omega \delta(\omega+\omega')(-1)^m\delta_{\ell,\ell'}\delta_{m,-m'},
\end{align}\es where the constants $c_{\ell,m;\ell',m';L,M}$ are Clebsch-Gordan coefficients. 
Note that the central charge $c$ can be different, which is consistent with the Jacobi identity.
We can further constrain $f$ or $h$. However, if $\dot h=0$, these two operators will be irrelevant and can be separated. 

}

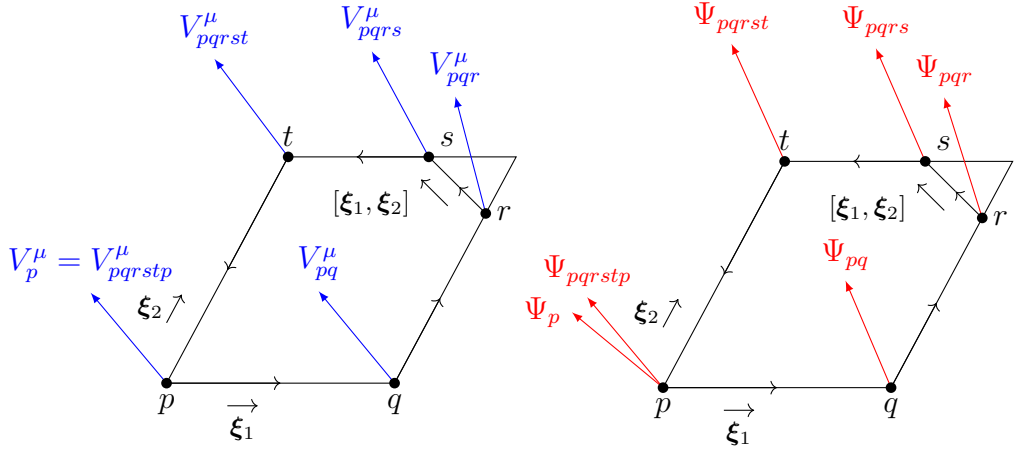
\begin{figure}
    \centering
    \begin{minipage}[h]{0.45\linewidth}
    \begin{tikzpicture}
      \draw (0,0) node[below] {$p$}--(3,0)node[below] {$q$}--(4.6,3)--(1.6,3) node[above] {$t$}-- cycle;
      \draw[->] (0,0)--(1.5,0);
      \draw[->] (0.8,-0.3)--(1.2,-0.3); 
      \node[below] at (1,-0.3) {\footnotesize$\bm\xi_1$};
      \draw[->] (3,0)--(3.6,1.125);
      \draw[->] (4.2,2.25) node[right] {$r$}--(3.85,2.6);
      \draw (3.85,2.6)--(3.45,3) node[above right] {$s$};
      \draw[->] (3.45,3)--(2.5,3);
      \draw[->] (1.6,3)--(0.8,1.5);
      \draw[->] (0,0.8)--(0.2,1.175); 
      \node[left] at (0.1,1.0) {\footnotesize$\bm\xi_2$};
      \draw[->] (3.7,2.35)--(3.35,2.7);
      \node[below left] at (3.35,2.7) {\footnotesize$[\bm\xi_1,\bm\xi_2]$};
      \draw[-latex,blue] (0,0)--(-1,1.2) node[above] {$V^\mu_p=V^\mu_{pqrstp}$};
      \draw[-latex,blue] (3,0)--(2,1.22) node[above] {$V^\mu_{pq}$};
      \draw[-latex,blue] (4.2,2.25)--(3.8,3.8) node[above] {$V^\mu_{pqr}$};
      \draw[-latex,blue] (3.45,3)--(2.7,4.4) node[above] {$V^\mu_{pqrs}$};
      \draw[-latex,blue] (1.6,3)--(0.63,4.3) node[above] {$V^\mu_{pqrst}$};
      \fill (0,0) circle (2pt);
      \fill (3,0) circle (2pt);
      \fill (1.6,3) circle (2pt);
      \fill (3.45,3) circle (2pt);
      \fill (4.2,2.25) circle (2pt);
    \end{tikzpicture}
  \end{minipage}\hspace*{-1cm}
  \begin{minipage}[h]{0.45\linewidth}
    \begin{tikzpicture}
      \draw (0,0) node[below] {$p$}--(3,0)node[below] {$q$}--(4.6,3)--(1.6,3) node[above] {$t$}-- cycle;
      \draw[->] (0,0)--(1.5,0);
      \draw[->] (0.8,-0.3)--(1.2,-0.3); 
      \node[below] at (1,-0.3) {\footnotesize$\bm\xi_1$};
      \draw[->] (3,0)--(3.6,1.125);
      \draw[->] (4.2,2.25) node[right] {$r$}--(3.85,2.6);
      \draw (3.85,2.6)--(3.45,3) node[above right] {$s$};
      \draw[->] (3.45,3)--(2.5,3);
      \draw[->] (1.6,3)--(0.8,1.5);
      \draw[->] (0,0.8)--(0.2,1.175); 
      \node[left] at (0.1,1.0) {\footnotesize$\bm\xi_2$};
      \draw[->] (3.7,2.35)--(3.35,2.7);
      \node[below left] at (3.35,2.7) {\footnotesize$[\bm\xi_1,\bm\xi_2]$};
      \draw[-latex,red] (0,0)--(-1,1.2) node[above] {$\Psi_{pqrstp}$};
      \draw[-latex,red] (0,0)--(-1.2,1) node[left] {$\Psi_p$};
      \draw[-latex,red] (3,0)--(2.4,1.42) node[above] {$\Psi_{pq}$};
      \draw[-latex,red] (4.2,2.25)--(3.7,3.85) node[above] {$\Psi_{pqr}$};
      \draw[-latex,red] (3.45,3)--(2.8,4.5) node[above] {$\Psi_{pqrs}$};
      \draw[-latex,red] (1.6,3)--(0.9,4.56) node[above] {$\Psi_{pqrst}$};
      \fill (0,0) circle (2pt);
      \fill (3,0) circle (2pt);
      \fill (1.6,3) circle (2pt);
      \fill (3.45,3) circle (2pt);
      \fill (4.2,2.25) circle (2pt);
    \end{tikzpicture}
  \end{minipage}
    \caption{In the left figure, a vector field is invariant under Lie transport around a loop. However, as shown in the right figure, a spinor field usually changes under Lie transport around the same loop.}
    \label{nonclosure}
\end{figure}

\paragraph{Non-closure of the Lie transport of the spinor field around a loop.}
The commutator between two superrotation generators \eqref{MYZ} is not closed due to the appearance of the helicity flux operator $\mathcal{O}_{h}$. This is phenomenon has been found in the bosonic theories where the  Lie derivative of the bulk tensor field $\bm T$ is always closed 
\be 
(\mathcal{L}_{\bm\xi_1}\mathcal{L}_{\bm\xi_2}-\mathcal{L}_{\bm\xi_2}\mathcal{L}_{\bm\xi_1})\bm T=\mathcal{L}_{[\bm\xi_1,\bm\xi_2]}\bm T
\ee and 
the non-closure of the commutator is from the introduction of the covariant variation, which modifies the bulk Lie derivative. However, in the fermionic case, the superrotation transformation of the boundary field $F$ could match with reduction from the Lie derivative of the bulk Dirac field $\Psi$. Therefore, there is a puzzle on the non-closure of the superrotation commutators. This is solved by noticing that the Lie derivative of the spinor field around a loop is actually not closed
\bea 
(\mathcal{L}_{\bm\xi_1}\mathcal{L}_{\bm\xi_2}-\mathcal{L}_{\bm\xi_2}\mathcal{L}_{\bm\xi_1})\Psi=\mathcal{L}_{[\bm\xi_1,\bm\xi_2]}\Psi+\frac{1}{16}(\partial_\rho\xi_1^\mu+\partial^\mu\xi_{1\rho})(\partial_\mu\xi_{2\sigma}+\partial_\sigma\xi_{2\mu})[\gamma^{\rho},\gamma^{\sigma}]\Psi.\label{anomalous}
\eea The above equation is independent of the parameter $\alpha$ and could be extended to curved spacetime. Note that the Lie derivative of the metric is 
\be 
\mathcal{L}_{\bm\xi}\eta_{\mu\nu}=\partial_\mu\xi_\nu+\partial_\nu\xi_\mu,
\ee and the Lie transport of the spinor field around a loop is only invariant when $\bm\xi$ is a conformal Killing vector. We may denote the second term in the right hand side of \eqref{anomalous} as
\bea 
\mathcal{A}(\bm\xi_1,\bm\xi_2,\Psi)=\frac{1}{16}\mathcal{L}_{\bm\xi_1}\eta_{\mu\rho}\mathcal{L}_{\bm\xi_2}\eta_{\nu\sigma}\eta^{\mu\nu}[\gamma^{\rho},\gamma^{\sigma}]\Psi.
\eea The non-closure of the Lie transport of the spinor field has already been noticed by \cite{Benn:1987fw,Fatibene:1996tf,2009arXiv0904.0258F} from other perspective. 
Note that \eqref{anomalous} should match with the commutator \eqref{MYZ} at the boundary. Therefore, we calculate the anomalous  term $\mathcal{A}$ near $\mathcal{I}^+$ 
\bs\begin{align}
    \mathcal{A}(\bm\xi_{f_1},\bm\xi_{f_2})&\dot{=}0,\\
    \mathcal{A}(\bm\xi_f,\bm\xi_{\bm Y})&\dot =0,\\
    \mathcal{A}(\bm\xi_{\bm Y},\bm\xi_{\bm Z})&\dot =-\frac{i}{2}o(\bm Y,\bm Z)\gamma^5\Psi^{(1)}.
\end{align}\es 
where $\dot{=}$ is to extract the $\mathcal{O}(r^{-1})$ term from the anomalous term. The above equations are consistent with the commutators \eqref{Tf1f2p},\eqref{TfMYp} and \eqref{MYZp}. We conclude that the helicity flux operator for the spinor field has already been hidden in the non-closure of the Lie transport of the spinor field. In figure \ref{nonclosure}, we have shown the difference between the Lie transport of the vector field and the spinor field around a closed loop.

\paragraph{Quantum flux operator from charge current.}{ Besides the chiral $U(1)_A$ symmetry, there is another $U(1)$ symmetry which leads to the charge current 
\be 
j_c^\mu=\bar\Psi\gamma^\mu\Psi.
\ee In a scattering process, massless Dirac particles radiate energy, angular momentum as well as electric charges to future null infinity. Therefore, one can find an associated electric charge flux density operator 
\be 
E=2:(\bar F F+G \bar G):
\ee and define the electric flux operator 
\bea 
\mathcal{E}_e=\int du d\Omega e(u,\Omega)E(u,\Omega)=2\int du d\Omega e(u,\Omega)(:\bar F F+G \bar G:).
\eea
{Similar to \eqref{fchiral}, we can derive the mode expansion of electric flux
\begin{align}
  \ce_{e=1}&=\int \frac{d^3\bm p}{(2\pi)^3 2\omega}(b_{-,\bm p}^\dagger b_{-,\bm p}-d_{+,\bm p}^\dagger d_{+,\bm p}+b_{+,\bm p}^\dagger b_{+,\bm p}-d_{-,\bm p}^\dagger d_{-,\bm p}),
\end{align}
which is the difference between numbers of particles and antiparticles, while \eqref{fchiral} is the difference between numbers of particles with opposite helicities.
}

One can add this operator into the algebra \eqref{spinoralgebrap}
\bs\begin{align}
    [\mathcal{T}_f,\mathcal{E}_e]&={\ce_{f\dot e}},\\
    [\mathcal{M}_{\bm Y},\mathcal{E}_e]&={i\mathcal{E}_{\bm Y(e)}},\\
    [\mathcal{O}_h,\mathcal{E}_e]&={0},\\
    [\mathcal{E}_{e_1},\mathcal{E}_{e_2}]&={ C_E(e_1,e_2)=\frac{ic}{2\pi}\mathcal{I}_{e_1\dot{e}_2-e_2\dot{e}_1}}.
\end{align}\es 
Similar to $C_O(h_1,h_2)$, $C_E(e_1,e_2)$ will also leads to a violation of the Jacobi identity. Since the electric flux operator is from an internal symmetry and not directly related to spacetime transformation {(namely not emerging from the commutator of generators for spacetime transformation like $\mathcal{O}_h$)}, we will not discuss it more in this paper.}

\section{Wess-Zumino model}\label{WZ}
In the previous section, we have derived the intertwined Carrollian diffeomorphism from the Dirac theory. Now we can extend the previous discussion to supersymmetric theories. To show the idea, we will consider the four-dimensional $\mathcal{N}=1$ supersymmetric theory found by Wess and Zumino \cite{Wess:1974tw}. The physical contents of the Wess-Zumino theory are a complex scalar field $\Phi(x)$ and a Majorana spinor $\Psi$ with the Lagrangian \cite{Martin:1997ns,Freedman:2012zz,Haber:2017aci}
\begin{align}
  \cl&=-\partial^\mu \bar{\Phi} \partial_\mu \Phi+\frac{i}{2}\bar{\Psi} \slashed \partial \Psi- g\bar\Psi(P_L\Phi+ P_R\bar\Phi)\Psi- g^2\Phi^2\bar\Phi^2 \label{L}
\end{align} where the projectors $P_{L/R}$ are
\be 
 P_{L/R}=\frac{1\mp\g_5}{2}.
\ee Note that we will consider the massless Wess-Zumino theory so that the mass term is absence in the Lagrangian. The cubic term from the scalar field also vanishes since the coupling constant is proportional to the mass. The Majorana condition 
\be 
\Psi=\Psi^C\equiv \cc \bar\Psi^{T},\quad \cc=
  \begin{pmatrix}
    \epsilon_{ab}&0\\ 0&\epsilon^{\dot a\dot b}
  \end{pmatrix}
\ee can be realized by a 
two-components Weyl spinor
\begin{align}
  \Psi=
  \begin{pmatrix}
    \chi_a \\ \chi^{\dagger\dot a}
  \end{pmatrix},\qquad
  \bar\Psi=(\chi^a,\chi^\dagger_{\dot a}).
\end{align} The Lagrangian can be written as 
\bea 
\mathcal{L}_{\text{sym}}=-\partial_\mu\Phi\partial^\mu\bar\Phi+\frac{i}{2}\chi\sigma^\mu\partial_\mu\chi^\dagger+\frac{i}{2}\chi^\dagger\bar\sigma^\mu\partial_\mu\chi-g(\Phi\chi\chi+\bar\Phi\chi^\dagger\chi^\dagger)-g^2\Phi^2\bar\Phi^2
\eea which is symmetric between $\chi$ and $\chi^\dagger$.
By throwing out a total derivative term, one can obtain
\begin{align}
  \cl&=-\partial^\mu \bar{\Phi} \partial_\mu \Phi+i\chi\s^\m\p_\m\chi^\dagger- g(\Phi\chi\chi+\bar\Phi\chi^\dagger\chi^\dagger)- g^2\Phi^2\bar\Phi^2. 
\end{align}
The SUSY transformation with parameter \(\varepsilon=(\eta_a,\eta^{\dagger\dot a})^T\) is \cite{Martin:1997ns}
\begin{subequations}\label{susytrans}
  \begin{align}
    \delta_\varepsilon \Phi(x)=\eta \chi, &\qquad \delta_\varepsilon \bar\Phi(x)=\eta^\dagger\chi^\dagger, \\
    \delta_\varepsilon \chi(x)=-i \s^\m\eta^\dagger\partial_\m \Phi - g\bar\Phi^2\eta, &\qquad \delta_\varepsilon \chi^\dagger(x)=i \eta\s^\m\partial_\m \bar\Phi -g\Phi^2\eta^\dagger. 
  \end{align}
\end{subequations}

\subsection{Equation of motion}
From the Lagrangian, one can read out the equation of motion
\begin{align}
  &\Box\Phi=g\chi^\dagger\chi^\dagger+2g^2\bar\Phi\Phi^2,\qquad \Box\bar\Phi=g\chi\chi+2g^2\Phi\bar\Phi^2,\label{eq1}\\
  &i\s^\m\p_\m\chi^\dagger=2g\Phi\chi,\qquad i\p_\m\chi\s^\m=-2g \bar\Phi\chi^\dagger\label{eq3}
\end{align} which can be solved  order by order with the fall-off conditions near $\mathcal{I}^+$
\be 
\Phi=\frac{\Sigma(u,\Omega)}{r}+\frac{\Sigma^{(2)}(u,\Omega)}{r^2}+\cdots,\qquad \chi=\frac{\psi(u,\Omega)}{r}+\frac{\psi^{(2)}(u,\Omega)}{r^2}+\cdots.
\ee Utilizing the identities
\begin{align}
  \p_\mu&=-n_\mu \partial_u+m_\mu\partial_r-\frac{1}{r}Y_\mu^A\partial_A,\\
  \partial_\mu\partial^\mu&=-2\partial_u\partial_r-\frac{2}{r}\partial_u+\partial_r^2+\frac{2}{r}\partial_r+\frac{1}{r^2}\nabla_A\nabla^A, \qquad \text{(on scalar)}
\end{align}
we find that the first equation of \eqref{eq1} gives
\begin{align}
  &\sum_{k=1}^\infty r^{-k-1}\big[(k-2)(k-1){\Sigma}^{(k-1)}+\nabla^2{\Sigma}^{(k-1)}+(2k-2)\dot{\Sigma}^{(k)}\big]\nn\\
  &=g\sum_{k=1}^\infty r^{-k-1}\sum_{m=1}^k \psi^{\dagger (m)}\psi^{\dagger(k+1-m)}+2g^2\sum_{k=2}^\infty r^{-k-1}\sum_{m+n\le k}\bar\Sigma^{(m)}\Sigma^{(n)}\Sigma^{(k+1-m-n)}.\label{bosonic}
\end{align}
At the leading \(r^{-2}\) order, the equation of motion is 
\be 
\psi^\dagger \psi^\dagger=0.\label{psipsi}
\ee To understand this equation, we consider the first equation of \eqref{eq3}
\begin{align}
  &-i\sigma^\mu\sum_{k=1}^\infty r^{-k}[n_\mu \dot \psi^{\dagger(k)}+(k-1)m_\mu\psi^{\dagger(k-1)}+Y^A_\mu\nabla_A\psi^{\dagger(k-1)}]=2 g\sum_{k=2}^\infty r^{-k}\sum_{m=1}^{k-1}\Sigma^{(m)}\psi^{(k-m)}.\label{fermionic}
\end{align}
At the \(r^{-1}\) order, we have
\begin{align}
  i\sigma^\mu n_\mu\dot\psi^\dagger=0 \quad\Rightarrow\quad \psi^{\dagger\dot a}(u, \Omega)=\lambda^{\dagger\dot a}\bar F(u,\Omega)+\varphi^{\dagger\dot a}(\Omega)
\end{align}
which solves \eqref{psipsi} for 
\be 
\varphi^{\dagger\dot a}(\Omega)\propto \lambda^{\dagger \dot a}.
\ee  The time independent term $\varphi^{\dagger\dot a}(\Omega)$ is not related to radiation, we may set it to zero from now on. Therefore, we could obtain two radiative modes $\Sigma$ and $F$. Both of them are free from equation of motion.
At the order $r^{-3}$, we have
\begin{align}
  \nabla^2{\Sigma}+2\dot{\Sigma}^{(2)}&=2g\psi^\dagger \psi^{\dagger(2)}+2g^2\bar\Sigma\Sigma^2\label{sigma2}
\end{align} from equation \eqref{bosonic}. Similaly, we find  
\bea 
-i\sigma^\mu[n_\mu\dot\psi^{(2)\dagger}+m_\mu\psi^\dagger+Y_\mu^A\nabla_A\psi^\dagger]=2g \Sigma\psi\label{psi2}
\eea at the order $r^{-2}$ from equation \eqref{fermionic}. We can solve $\psi^{(2)}$ after imposing the initial condition (the static mode below) at the time $u=u_0$
\bea 
\psi^{(2)\dagger\dot a}=\lambda^{\dagger\dot a}\bar F^{(2)}+\kappa^{\dagger\dot a}\bar{\tilde{F}}^{(2)}+\text{static mode}
\eea with 
\bea 
\dot{\bar{\tilde{F}}}^{(2)}=-\frac{1}{2}(\bar\zeta^A\nabla_A\bar F+\frac{1}{2}\nabla_A\bar\zeta^A \bar F)+ig\Sigma F.
\eea The Hermitian conjugate of the above solution is 
\bea 
\psi^{(2)}_a=\lambda_a F^{(2)}+\kappa_a \tilde{F}^{(2)}+\text{static mode}
\eea 
 with 
\bea 
{\dot{\tilde{{F}}}^{(2)}=-\frac{1}{2}(\zeta^A \nabla_A F+\frac{1}{2} \nabla_A \zeta^A F)-ig\bar\Sigma\bar F}.
\eea 
Substituting the solution into the equation \eqref{sigma2}, we can solve $\Sigma^{(2)}$  up to an initial data. In summary, the above coupling equations can be solved iteratively. In this work, we only need the solution up to $\psi^{(2)}$. 

\subsection{Fluxes}
Under \(x^\mu \rightarrow x^\mu+\epsilon^\mu\), the complex scalar and spinor field have the same transformation law
\begin{align}
  \delta \Phi(x)=-\epsilon^\mu \partial_\mu \Phi(x),\qquad \delta {\chi}(x)=-\epsilon^\mu \partial_\mu {\chi}(x),
\end{align}
and the Lagrangian changes as \(\delta \mathcal{L}(x)=\partial_\mu\left[-\epsilon^\mu \mathcal{L}(x)\right]\). Using Noether's theorem
\begin{align}
  {J}_\epsilon^\mu & =-\epsilon^\nu \partial_\nu \Phi \frac{\partial \mathcal{L}}{\partial\left(\partial_\mu \Phi\right)}-\epsilon^\nu \partial_\nu \bar{\Phi} \frac{\partial \mathcal{L}}{\partial\left(\partial_\mu \bar{\Phi}\right)}-\epsilon^\nu \partial_\nu \chi \frac{\partial \mathcal{L}}{\partial\left(\partial_\mu \chi\right)}-\epsilon^\nu \frac{\partial \mathcal{L}}{\partial\left(\partial_\mu {\chi}^\dagger\right)}\partial_\nu {\chi}^\dagger +\epsilon^\mu \mathcal{L} ,
\end{align}
we obtain a stress tensor from \({J}_\epsilon^\mu=\epsilon_\nu {T}^{\mu \nu}\)
\begin{align}
  {T}^{\mu \nu}&=\partial^\mu\bar\Phi\partial^\nu\Phi+\partial^\nu\bar\Phi\partial^\mu\Phi{ -}\frac{i}{4}(\chi^\dagger\bar\sigma^\mu\partial^\nu\chi+\chi^\dagger\bar\sigma^\nu\partial^\mu\chi+\chi\sigma^\mu\partial^\nu\chi^\dagger+\chi\sigma^\nu\partial^\mu\chi^\dagger)+\eta^{\mu\nu}\mathcal{L}_{\text{sym}}
\end{align} where we have used the symmetric Lagrangian.

To obtain the supercurrent,
we compute the variation of the symmetric Lagrangian under supersymmetric transformation \eqref{susytrans}
\bea 
\delta_\epsilon\mathcal{L}_{\text{sym}}=\frac{1}{2}\partial_\mu(\eta^\dagger\bar\sigma^\mu\sigma^\nu\chi^\dagger\partial_\nu\Phi+\eta\sigma^\mu\bar\sigma^\nu\chi\partial_\nu\bar\Phi+ig\chi^\dagger\sigma^\mu\eta\bar\Phi^2+ig\chi\sigma^\mu\eta^\dagger\Phi^2).
\eea On the other hand, the variation of the action leads to 
\bea 
\delta S=\delta \int d^4 x \mathcal{L}_{\text{sym}}=({\rm EOM})+\int d^4 x\partial_\mu\Theta^\mu
\eea with 
\be 
\Theta^\mu=-\partial^\mu\Phi\delta\bar\Phi-\partial^\mu\bar\Phi \delta\Phi+\frac{i}{2}\chi\sigma^\mu\delta\chi^\dagger+\frac{i}{2}\chi^\dagger\bar\sigma^\mu\delta\chi.
\ee Therefore, we find the supercurrent  
\bea 
J_{\text{susy}}^\mu&=&\frac{1}{2}(\eta^\dagger\bar\sigma^\mu\sigma^\nu\chi^\dagger\partial_\nu\Phi+\eta\sigma^\mu\bar\sigma^\nu\chi\partial_\nu\bar\Phi+ig\chi^\dagger\sigma^\mu\eta\bar\Phi^2+ig\chi\sigma^\mu\eta^\dagger\Phi^2)-\Theta^\mu\nn\\&=&\partial^\mu\Phi\eta^\dagger\chi^\dagger+\partial^\mu\bar\Phi\eta\chi+\frac{1}{2}\eta\sigma^{\mu\nu}\chi\partial_\nu\bar\Phi+\frac{1}{2}\eta^\dagger\bar\sigma^{\mu\nu}\chi^\dagger\partial_\nu\Phi+ig\Phi^2 \chi\sigma^\mu\eta^\dagger+ig\bar\Phi^2\chi^\dagger\sigma^\mu\eta\nn\\
\eea where
\bea 
\sigma^{\mu\nu}=\sigma^\mu\bar\sigma^\nu -\sigma^\nu\bar\sigma^\mu,\quad \bar\sigma^{\mu\nu}=\bar\sigma^\mu\sigma^\nu-\bar\sigma^\nu\sigma^\mu.
\eea 
By computing the fluxes across $\mathcal{I}^+$, we may define the following quantum flux operators
\begin{align}
  &\ct_f=\ct_f^{\rm f}+\ct_f^{\rm b}=\int dud\Omega f(u,\Omega)[T^{\rm f}(u,\Omega)+T^{\rm b}(u,\Omega)],\\ 
  &\cm_{\bm Y}=\cm_{\bm Y}^{\rm f}+\cm_{\bm Y}^{\rm b}=\int dud\Omega Y^A(\Omega)[M_A^{\rm f}(u,\Omega)+M_A^{\rm b}(u,\Omega)]-\frac{i}{8}\co_{o(y, \bar{y})},\\
  &\co_h=\int d u d \Omega\, h(u,\Omega)O^{\rm f}(u,\Omega),\\
  &\cq_{\eta}=\int dud\Omega\,\eta(u,\Omega)Q(u,\Omega),\qquad \bar\cq_{\bar\eta}=\int dud\Omega\,\bar\eta(u,\Omega)\bar Q(u,\Omega),
\end{align}
where the flux densities read
\begin{align}
  &T^{\rm f}(u,\Omega)= i(:\bar{F} \dot{F}-\dot{\bar{F}} F:),\qquad T^{\rm b}(u,\Omega)=2:\dot\Sigma\dot{\bar\Sigma}:\,=\,:\dot\Sigma\dot{\bar\Sigma}+\dot{\bar\Sigma}\dot\Sigma:,\\
  &M_A^{\rm f}(u,\Omega)=i\left(:\bar{F} \nabla_A F-\nabla_A \bar{F} F:\right),\\
  &M_A^{\rm b}(u,\Omega)=\frac{1}{2}(:\dot\Sigma\nabla_A\bar\Sigma+\dot{\bar\Sigma}\nabla_A\Sigma-\Sigma\nabla_A\dot{\bar\Sigma}-{\bar\Sigma}\nabla_A\dot\Sigma:),\\
  &O^{\rm f}(u,\Omega)=\,2:\bar FF:,\\
  &Q(u,\Omega)=2\,:\dot{\bar\Sigma}F:,\qquad \bar Q(u,\Omega)=2\,:\dot{\Sigma}\bar F:.
\end{align}
The energy  flux operator is exactly the summation of the energy flux operators from the complex scalar and the Majorana fermion. We have used the superscript $b/f$ to denote their contributions correspondingly.

\paragraph{Fluxes acting on fields.}
The non-vanishing fundamental commutators/anti-commutators can be read out from the scalar/fermionic theory
\bea 
[\Sigma(u,\Omega),\bar\Sigma(u',\Omega')]=\frac{i}{2}\alpha(u-u')\delta(\Omega-\Omega'),\\ \{F(u,\Omega),\bar F(u',\Omega')\}=\frac{1}{2}\delta(u-u')\delta(\Omega-\Omega').
\eea Therefore, the energy flux operator acts on the boundary fields as 
\bs\begin{align}
&[\mathcal{T}_f,\Sigma]=-if\dot\Sigma,\quad [\mathcal{T}_f,\bar\Sigma]=-if\dot{\bar\Sigma},\\
&[\mathcal{T}_f,F]=-if\dot F-\frac{i}{2}\dot f F,\quad [\mathcal{T}_f,\bar F]=-if\dot{\bar F}-\frac{i}{2}\dot f\bar F
\end{align}\es which agrees with the bulk diffeomorphism generated by $\bm\xi_f$. The angular momentun flux is also the summation of two parts and it acts on the fundamental fields as 
\bs\begin{align}
    &[\mathcal{M}_{\bm Y},\Sigma]=-iY^A\nabla_A\Sigma-\frac{i}{2}\nabla\cdot Y \Sigma,\\ &[\mathcal{M}_{\bm Y},\bar\Sigma]=-iY^A\nabla_A\bar \Sigma-\frac{i}{2}\nabla\cdot Y \bar\Sigma,\\
    &[\mathcal{M}_{\bm Y},F]=-iY^A\nabla_AF-\frac{i}{2}\nabla\cdot Y F+\frac{i}{8}o(y,\bar y) F,\\ &[\mathcal{M}_{\bm Y},\bar F]=-iY^A\nabla_A\bar F-\frac{i}{2}\nabla\cdot Y \bar F-\frac{i}{8}o(y,\bar y) \bar F.
\end{align}\es which matches with the bulk diffeomorphism generated by $\bm\xi_{\bm Y}$ (after subtracting a part which is generated by $\bm\xi_{f=\frac{1}{2}u\nabla\cdot Y}$). The helicity flux operator $\mathcal{O}_{h}$ only depends on the fermionic boundary field as there is no chiral symmetry for the scalar field. 

The helicity flux operator acts on the fundamental fields as 
\begin{align}
    &[\mathcal{O}_{h},\Sigma(u,\Omega)]=[\mathcal{O}_h,\bar\Sigma(u,\Omega)]=0,\label{helicitysigma}\\ &[\mathcal{O}_h,F(u,\Omega)]=-h(u,\Omega)F(u,\Omega),\quad [\mathcal{O}_h,\bar{F}(u,\Omega)]=h(u,\Omega)\bar F(u,\Omega),
\end{align} which agrees with the bulk chiral transformation
\bea 
\Psi\to e^{ih\gamma_5}\Psi\quad\Rightarrow\quad \delta_h\chi=-ih\chi,\quad \delta_h\chi^\dagger=ih\chi^\dagger\quad\Rightarrow\quad \delta_h F=-ih F,\quad \delta_h\bar F=ih\bar F \label{chiralwz}
\eea for constant $h$. Note that \eqref{chiralwz} alone is not the symmetry transformation of Wess-Zumino model. Actually, for the action to be invariant under the transformation \eqref{chiralwz}, one should transform the field $\Phi$ as follows
\be 
\Phi\to e^{2ih}\Phi\label{3.48}
\ee to cancel the phase in the Yukawa coupling. However, the above transformation law for the scalar field does not match with \eqref{helicitysigma}. {Actually, the phase transformation of complex scalar can give a (electric) charge flux 
\begin{align}
  \ce^{\rm b}_e=\int d u d \Omega\, e(\Omega) [-2i(:\bar\Sigma\dot\Sigma-\Sigma\dot{\bar\Sigma}:)],\label{3.49}
\end{align}
and one can immediately work out
\begin{align}
    [\mathcal{E}^{\rm b}_e,\Sigma(u,\Omega)]=2e(\Omega)\Sigma(u,\Omega),\qquad [\mathcal{E}^{\rm b}_e,\bar\Sigma(u,\Omega)]=-2e(\Omega)\bar\Sigma(u,\Omega),
\end{align}
which agrees with \eqref{3.48}. We do not include this flux since it can not emerge from the commutator of superrotation generators, and it is not appropriate to say spin 0 field can form a helicity flux whose meaning is the difference between numbers of particles with opposite helicities. However, one can easily find the mode expansion of this flux. Suppose the mode expansion of the bulk complex scalar \begin{align}
  \Phi(x)=\int \frac{d^3 \bm p}{(2 \pi)^3 2 \omega}[a_{\bm p} e^{i p \cdot x}+c_{\bm p}^{\dagger}e^{-i p \cdot x}],
\end{align} 
which reduces to $\ci^+$ as
\begin{align}
  \Sigma(u, \Omega)=\int \frac{d \omega }{8 \pi^2 i}[a_{\omega}(\Omega) e^{-i \omega u} -c_{\omega}^{\dagger}(\Omega)  e^{i \omega u}].
\end{align}
Then we can compute 
\begin{align}
  \ce^{\rm b}_{e=1}&=-2i\int dud\Omega(:\bar\Sigma\dot\Sigma-\Sigma\dot{\bar\Sigma}:)=-2\int \frac{d^3\bm p}{(2\pi)^32\omega}[a^{\dagger}_{\bm p}a_{\bm p}-c^{\dagger}_{\bm p}c_{\bm p}].
\end{align}
This is also a difference between numbers of two kinds of particles, not between opposite helicities but between the particles and their antiparticles. We will discuss this point at the end of subsection \ref{sec3.4}.}

The superflux operator  $Q_\eta,\bar Q_{\bar\eta}$ acts on the fundamental fields as
\begin{align}
  [\cq_{\eta},\Sigma]=-i\eta F,\quad [\bar\cq_{\bar\eta},\bar\Sigma]=-i\bar\eta\bar F; \quad [\cq_{\eta},\bar F]=\eta\dot{\bar\Sigma},\quad [\bar\cq_{\bar\eta},F]=\bar\eta\dot{\Sigma}
\end{align} which 
agrees with the bulk reduction of SUSY transformation 
\begin{align}
\delta_\varepsilon \Sigma=\eta^a\l_a F, &\qquad \delta_\varepsilon \bar\Sigma=\eta^\dagger_{\dot a}\l^{\dagger\dot a}\bar F, \\
  \delta_\varepsilon F=-i \l^\dagger_{\dot a}\eta^{\dagger\dot a}\dot\Sigma, &\qquad \delta_\varepsilon\bar F=i \eta^a\l_{a}\dot{\bar\Sigma}
\end{align} for constant Grassmannian $\eta^a$. Note that we have chosen the twistor $\lambda_a$ commutes with the Grassmannian. The quantity $\eta$ in the superflux operator is a one-component Grassmannian.

\subsection{Supersymmetric intertwined Carrollian diffeomorphism}
Now we can compute the supersymmetric version of the intertwined Carrollian diffeomorphism. 
\begin{subequations}\label{susyicd}
  \begin{align}
    &[\mathcal{T}_{f_1},\mathcal{T}_{f_2}]=C_T(f_1,f_2)+i\mathcal{T}_{f_1\dot{f}_2-f_2\dot{f_1}},&&
    [\mathcal{T}_{f},\mathcal{M}_{\bm Y}]=-i\mathcal{T}_{\bm Y(f)},\\
    &[\mathcal{T}_f,\mathcal{O}_h]=i\mathcal{O}_{f\dot h},&&
    [\mathcal{M}_{\bm Y},\mathcal{M}_{\bm Z}]=i\mathcal{M}_{[\bm Y, \bm Z]}-\frac{i}{2}\co_{o(\bm Y, \bm Z)},\\
    &[\mathcal{M}_{\bm Y},\mathcal{O}_h]=i\mathcal{O}_{\bm Y(h)},&&
    [\mathcal{O}_{h_1},\mathcal{O}_{h_2}]=C_{O}(h_1,h_2),\\
    &[\ct_f,\cq_{\eta}] =i\cq_{f\dot\eta-\frac{1}{2}\dot f\eta},&&  [\ct_f,\bar\cq_{\bar\eta}]=i\bar\cq_{f\dot{\bar\eta}-\frac{1}{2}\dot f\bar\eta},\\
    &[\cm_{\bm Y},\cq_\eta]=i\cq_{Y^A\nabla_A\eta+\frac{1}{8}o(y,\by)\eta},&& [\cm_{\bm Y},\bar\cq_{\bar\eta}]=i\bar\cq_{Y^A\nabla_A\bar\eta-\frac{1}{8}o(y,\by)\bar\eta},\\
    &[\co_h,\cq_\eta]=-\cq_{h\eta},&& [\co_h,\bar\cq_{\bar\eta}]=\bar\cq_{h\bar\eta},\\
    &[\cq_{\eta_1},\bar\cq_{\bar\eta_2}]=C_Q(\eta_1, \bar\eta_2)-\ct_{\eta_1\bar\eta_2}-\frac{i}{2}\co_{\dot\eta_1 \bar\eta_2+\dot{\bar\eta}_2\eta_1},&& [\cq_{\eta_1},\cq_{\eta_2}]=[\bar\cq_{\bar\eta_1},\bar\cq_{\bar\eta_2}]=0.\label{qq}
  \end{align}
\end{subequations} The superrotation generator $\mathcal{M}_{\bm Y}$ has a contribution from the complex scalar and then one should turn off the time-dependent of $\bm Y$ to avoid the non-local terms. The helicity flux operator is only constructed by fermionic field and then we can still allow the time-dependence of $h$. Interestingly, the SUSY parameters $\eta$ can also be Grassmannians on $\mathcal{I}^+$, which is time-dependent in general. 
The central charges are
\bs\begin{align}
C_T(f_1,f_2)&=-\frac{ic}{16\pi}\mathcal{I}_{f_1\dddot{f}_2-f_2\dddot{f}_1},\label{ct}\\
C_{O}(h_1,h_2)&=\frac{ic}{4\pi}\mathcal{I}_{h_1\dot{h}_2-h_2\dot{h}_1},\label{co}\\
C_Q(\eta_1,\bar\eta_2)&=\frac{c}{8\pi}\mathcal{I}_{\ddot{\eta}_1\bar\eta_2+\eta_1\ddot{\bar\eta}_2}.\label{cq}
\end{align}\es 

Note that the coefficient of the central charge $C_T$ is $3/2$ times the one in the complex scalar theory. This is because the Majorana fermion has only half of degrees of freedom compared with the Dirac fermion while the latter contributes to the same central charge as a complex scalar.  The central charges \eqref{co} is half of the one in the Dirac theory since only Majorana fermion contributes to this number. Besides the central charges that have been known in Dirac theory, we also find a central charge from the commutator of the (generalized) superfluxes. This central charge would be zero once $\eta$ is time-independent. 

\paragraph{Helicity flux operator from SUSY.}
We are familiar with the emergence of the helicity flux operator from the commutator of the angular momentum flux operators. In the SUSY extension, we notice that the commutator of the (generalized) superfluxes are composed with three parts. The first part is the central charge and the second part is the energy flux operator $\mathcal{T}_{\eta_1\bar\eta_2}$ whose appearance is expected since the commutator of the (global) superfluxes is proportional to the four-momentum operator in supersymmetric theories. The last term is the helicity flux operator which only appeared for time-dependent test functions. This is a new feature which has never been noticed in the literature. The helicity flux operator characterizes the time and angular distribution of the intrinsic helicity flux  of the spinning field. As the superflux connects the particles with different spins, it may not be a surprise that 
the helicity flux appears in the right hand side of the commutator of the (generalized) superfluxes.

{
\paragraph{Jacobi identities.} We find that the nontrivial central charges \eqref{ct}-\eqref{cq} will lead to the violation of Jacobi identities for the flux operator triples $\ct\ct\cm$, $\co\co\cm$ and $\cq\bar\cq\cm$
\begin{subequations}
  \begin{align}
  J[\ct_{f_1},\ct_{f_2},\cm_{\bm Y}]&=-\frac{c}{16\pi}\ci_{Y^A\nabla_A(f_1\p_u^3f_2-\p_u^3f_1f_2)},\\ 
  J[\mathcal{O}_{h_1},\mathcal{O}_{h_2},\mathcal{M}_{\bm Y}]&=\frac{c}{4\pi}\ci_{Y^A\nabla_A(h_1\dot h_2-h_2\dot h_1)},\\ 
  J[\cq_{\eta_1},\bar\cq_{\bar\eta_2},\mathcal{M}_{\bm Y}]&=-\frac{ic}{8\pi}\ci_{Y^A\nabla_A(\ddot\eta_1\bar\eta_2+\eta_1\ddot{\bar\eta}_2)},
\end{align}
\end{subequations}
where the first two is similar to the Dirac theory.


To satisfy the Jacobi identity, we should take  \(f(u,\Omega)=T(\Omega)+u W(\Omega)+u^2 V(\Omega)\), $h=h(\Omega)$,  \(\eta(u,\Omega)=u\m(\Omega)+\n(\Omega)\) and \(\bar\eta(u,\Omega)=u\bar\m(\Omega)+\bar\n(\Omega)\)
  \begin{subequations}\label{lwzal}
    \begin{align}
      &[\mathcal{T}_{f_1},\mathcal{T}_{f_2}]=i\mathcal{T}_{f_1\dot f_2-f_2\dot f_1},&&
      [\mathcal{T}_{f},\mathcal{M}_{\bm Y}]=-i\mathcal{T}_{\bm Y(f)},\\
      &[\mathcal{T}_f,\mathcal{O}_h]=0,&&
      [\mathcal{M}_{\bm Y},\mathcal{M}_{\bm Z}]=i\mathcal{M}_{[\bm Y, \bm Z]}-\frac{i}{2}\co_{o(\bm Y, \bm Z)},\\
      &[\mathcal{M}_{\bm Y},\mathcal{O}_h]=i\mathcal{O}_{\bm Y(h)},&&
      [\mathcal{O}_{h_1},\mathcal{O}_{h_2}]=0,\\
      &[\ct_f,\cq_{\eta}] = { i\mathcal Q_{f\dot\eta-\frac{1}{2}\dot f \eta}},&&  [\ct_f,\bar\cq_{\bar\eta}]= i\bar{\mathcal Q}_{f\dot{\bar\eta}-\frac{1}{2}\dot f\bar \eta}\\
      &[\cm_{\bm Y},\cq_\eta]=i\cq_{Y^A\nabla_A\eta+\frac{1}{8}o(y,\by)\eta},&& [\cm_{\bm Y},\bar\cq_{\bar\eta}]=i\bar\cq_{Y^A\nabla_A\bar\eta-\frac{1}{8}o(y,\by)\bar\eta},\\
      &[\co_h,\cq_\eta]=-\cq_{h\eta},&& [\co_h,\bar\cq_{\bar\eta}]=\bar\cq_{h\bar\eta},\\
      &[\cq_{\eta_1},\bar\cq_{\bar\eta_2}]=-\ct_{\eta_1\bar\eta_2}-\frac{i}{2}\co_{\dot\eta_1 \bar\eta_2+\dot{\bar\eta}_2\eta_1},&& [\cq_{\eta_1},\cq_{\eta_2}]=[\bar\cq_{\bar\eta_1},\bar\cq_{\bar\eta_2}]=0.\label{QQbT}
    \end{align}
  \end{subequations}
To be more explicit, the right-hand side of $[\cq_{\eta_1},\bar\cq_{\bar\eta_2}]$ reads
\begin{align}
    -\ct_{u^2\mu_1\bar\m_2+u(\m_1\bar\nu_2+\n_1\bar\mu_2)+\n_1\bar\n_2}-\frac{i}{2}\co_{\m_1\bar\n_2+\bar\m_2\n_1}
\end{align}
which will not result in a linear time-dependence for the parameter of helicity flux.
  
One can further require $f=T(\Omega)+uW(\Omega)$ as well as $\dot\eta=\dot{\bar\eta}=\dot h=0$ which will leads to a supersymmetric BMSW algebra intertwined with the helicity flux (we only list the lines with changes)
\begin{subequations}
  \begin{align}
    &[\mathcal{T}_{f_1},\mathcal{T}_{f_2}]=i\mathcal{T}_{T_1W_2-T_2W_1},&&
    [\mathcal{T}_{f},\mathcal{M}_{\bm Y}]=-i\mathcal{T}_{\bm Y(f)},\\
    &[\ct_f,\cq_{\eta}] =-\frac{1}{2}i\cq_{W\eta},&&  [\ct_f,\bar\cq_{\bar\eta}]=-\frac{1}{2}i\bar\cq_{W\bar\eta},\\
    &[\cq_{\eta_1},\bar\cq_{\bar\eta_2}]=-\ct_{\eta_1\bar\eta_2},&& [\cq_{\eta_1},\cq_{\eta_2}]=[\bar\cq_{\bar\eta_1},\bar\cq_{\bar\eta_2}]=0.
  \end{align}
\end{subequations}

We can further require $\dot f=0$ which implies $\dot\eta=\dot{\bar\eta}=\dot h=0$ since the time-dependence of $\eta,\bar\eta$ comes from $[\ct,\cq]$ and $[\ct,\bar\cq]$ while the time-dependence of $h$ comes from $[\cq,\bar\cq]$. These constraints will significantly reduce the algebra to
\begin{subequations}
  \begin{align}
    &[\mathcal{T}_{f_1},\mathcal{T}_{f_2}]=0,&&
    [\mathcal{T}_{f},\mathcal{M}_{\bm Y}]=-i\mathcal{T}_{\bm Y(f)},\\
    &[\ct_f,\cq_{\eta}] =0,&&  [\ct_f,\bar\cq_{\bar\eta}]=0,\\
    &[\cq_{\eta_1},\bar\cq_{\bar\eta_2}]=-\ct_{\eta_1\bar\eta_2},&& [\cq_{\eta_1},\cq_{\eta_2}]=[\bar\cq_{\bar\eta_1},\bar\cq_{\bar\eta_2}]=0.
  \end{align}
\end{subequations}
This is a supersymmetric extension of the generalized BMS algebra intertwined with the helicity flux.

Similar to the Dirac theory, one may also require $Y^A=\epsilon^{AB}\nabla_B \mathcal Y$ to rule out Lorentz boost but save the Jacobi identity with non-trivial central charges. 
Another way is to exclude $\cm_{\bm Y}$, we can get a closed Lie algebra and $\co_h$ can appear from superfluxes this time (not like in the Dirac theory, only superrotations can produce helicity flux). However, for these alternative choices, the resulting algebras do not include the Poincar\'e algebra as a subalgebra.
}

\subsection{Reducing to well-known algebras}\label{sec3.4}
\paragraph{Comparison with $\mathcal{N}=1$ supersymmetric BMS algebra.} { We have compared our flux algebra with the super-BMS algebra in \cite{Fotopoulos:2020bqj}. It turns out that if we identify two conventions as follows
\begin{align}
  P_{kl}\leftrightarrow \ct_{f_{kl}},\quad L_m\leftrightarrow \cm_{{\bm Y}_m}+{ \mathcal{T}_{\frac{1}{2}u\nabla\cdot Y_m}},\quad \bar L_m\leftrightarrow \cm_{\bar{\bm Y}_m}+{ \mathcal{T}_{\frac{1}{2}u\nabla\cdot {\bar Y}^{}_m}},\quad G_n\leftrightarrow \cq_{\eta_n},\quad \bar G_n\leftrightarrow \bar\cq_{\bar\eta_n},\label{3.55}
\end{align}
where the parameters take \footnote{Note that $\eta,\bar\eta$ taking $\l_a,\l^{\dagger\dot a}$ means the parameters become usual numbers rather than Grassmann numbers which needs careful handling. One should change the commutators involving two $\mathcal{Q}/\bar{\mathcal Q}$ to anti-commutators. There could also be sign flipping in \eqref{QQbT}. Moreover, $\bm Y_m=Y_m^z\partial_z$ and $\bar{\bm Y}_m=\bar Y_m^{\bar z}\partial_{\bar z}$ are independent vectors. }
\begin{align}
  f_{kl}=z^{1/2-k}\bz^{1/2-l},\quad Y^z_m=-iz^{1-m},\quad \bar Y^{\bz}_m=i\bar z^{1-m}, \quad \eta_n=z^{1/2-n},\quad \bar\eta_n=\bar z^{1/2-n},
\end{align}
and we should replace the spherical metric with the one for the complex plane\footnote{{In our formalism, we always use metric and coordinates on the unit sphere. However, to match with many existing results in celestial holography, one should turn to the complex plane. To be more precise, }this implies that
\begin{align}
    \g_{z\bz}=\z_{(z}\bar\z_{\bz)}=1,\quad \bar\z_{\bz}=\z_z={ \z^{\bz}=\bar\z^z}=\sqrt2,\quad \bar\z_z=\z_{\bz}={\z^{z}=\bar\z^{\bar z}}=0,\quad \epsilon^{z\bar z}=i=-\epsilon^{\bar z z}.\nn
\end{align}
The definition of $\gamma_{AB}$ and $\epsilon_{AB}$ are consistent with identities \eqref{zetaidentity}.
}. We will find
\begin{subequations}
  \begin{align}
  &[\ct_{f_{kl}},\ct_{f_{mn}}]=0,\qquad [\cm_{Y_n}+\ct_{\frac{1}{2}u\nabla\cdot Y_n},\ct_{f_{kl}}]=(\frac{1}{2}n-k)\ct_{f_{k+n,l}},\\
  &[\cm_{Y_m}+\ct_{\frac{1}{2}u\nabla\cdot Y_m},\cm_{Y_n}+\ct_{\frac{1}{2}u\nabla\cdot Y_n}]=(m-n)(\cm_{Y_{m+n}}+\ct_{\frac{1}{2}u\nabla\cdot Y_{m+n}}),\\
  &[\cm_{\bar Y_m}+\ct_{\frac{1}{2}u\nabla\cdot\bar Y_m},\cm_{\bar Y_n}+\ct_{\frac{1}{2}u\nabla\cdot\bar Y_n}]=(m-n)(\cm_{\bar Y_{m+n}}+\ct_{\frac{1}{2}u\nabla\cdot\bar Y_{m+n}}),\\
  &[\cm_{Y_m}+\ct_{\frac{1}{2}u\nabla\cdot Y_m},\cm_{\bar Y_n}+\ct_{\frac{1}{2}u\nabla\cdot\bar  Y_n}]=0,\qquad \{\cq_{\eta_m},\bar\cq_{\bar\eta_n}\}=\ct_{f_{mn}},\\
  &[\ct_{f_{kl}},\cq_{\eta_n}]=[\ct_{f_{kl}},\bar\cq_{\bar\eta_n}]=0, \qquad \{\cq_{\eta_m},\cq_{\eta_n}\}=\{\bar\cq_{\bar\eta_m},\bar\cq_{\bar\eta_n}\}=0,\\
  &[\cm_{Y_m}+\ct_{\frac{1}{2}u\nabla\cdot Y_m},\cq_{\eta_n}]=(\frac{m}{2}-n)\cq_{\eta_{m+n}},\\ 
  &[\cm_{\bar Y_m}+\ct_{\frac{1}{2}u\nabla\cdot \bar Y_m},\bar\cq_{\bar\eta_n}]=(\frac{m}{2}-n)\bar\cq_{\bar\eta_{m+n}},
\end{align}
\end{subequations}
which match with the results in \cite{Fotopoulos:2020bqj}. 
}

{\paragraph{Including the helicity operator.} If we adopt the identification \eqref{3.55}, the helicity flux operator can not be naturally introduced since the superrotation and superflux parameters are split into holomorphic and anti-holomorphic parts. As a result, $[\co,\cq]$ and $[\co,\bar\cq]$ can not both have a nice form. Therefore, we take a more general parameterization
\begin{align}
  f_{kl}=z^{-k}\bz^{-l},\quad {\tt y}_{mn}=\sqrt2z^{-m}\bz^{-n}=\bar{\tt y}_{nm},\quad \eta_{kl}=z^{-k}\bz^{-l}=\bar\eta_{lk},\quad h_{mn}=z^{-m}\bz^{-n},
\end{align}
where ${\tt y}$ and $\bar{\tt y}$ are related to $Y^A$ through
\begin{align}
    Y^z=\frac{1}{2}{\tt y}\bar\z^z=\frac{1}{\sqrt2}{\tt y},\qquad Y^{\bz}={ \frac{1}{2}\bar{\tt y}\z^{\bar z}=\frac{1}{\sqrt2}\bar{\tt y}}.
\end{align}
Then we have the following commutators
\begin{subequations}
  \begin{align}
  & [\ct_{f_{kl}},\ct_{f_{mn}}]=[\ct_{f_{kl}},\co_{h_{mn}}]=[\co_{h_{kl}},\co_{h_{mn}}]=0,\\
  &[\ct_{f_{kl}},\cq_{\eta_{mn}}]=[\ct_{f_{kl}},\bar\cq_{\bar\eta_{mn}}]= \{\cq_{\eta_{mn}},\cq_{\eta_{pq}}\}=\{\bar\cq_{\bar\eta_{mn}},\bar\cq_{\bar\eta_{pq}}\}=0,\\
  &\{\cq_{\eta_{mn}},\bar\cq_{\bar\eta_{pq}}\}=\ct_{f_{m+q,n+p}}\\ &[\co_{h_{mn}},\cq_{\eta_{kl}}]=-\cq_{\eta_{m+k,n+l}},\quad [\co_{h_{mn}},\bar\cq_{\bar\eta_{kl}}]=-\bar\cq_{\eta_{m+l,n+k}},\\
  &[\cm_{Y_{mn}},\ct_{f_{kl}}]=-ik\ct_{f_{m+k+1,n+l}}-il\ct_{f_{n+k,m+l+1}},\\
  &[\cm_{Y_{mn}},\co_{h_{pq}}]=-ip\co_{h_{m+p+1,n+q}}-iq\co_{h_{n+p,m+q+1}},\\
  &[\cm_{Y_{mn}},\cq_{\eta_{pq}}] =i(\frac{1}{4}m-p)\cq_{\eta_{m+p+1,n+q}}-i(\frac{1}{4}m+q)\cq_{\eta_{n+p,m+q+1}},\\
  &
[\cm_{Y_{mn}},\bar\cq_{\bar\eta_{pq}}]=i(\frac{1}{4}m-p)\bar\cq_{\bar{\eta}_{m+p+1,n+q}}-i(\frac{1}{4}m+q)\bar\cq_{\bar{\eta}_{n+p,m+q+1}},
\end{align}
\end{subequations}

and the last one
{
\begin{align}
    [\cm_{Y_{mn}},\cm_{Y_{pq}}]&=i(m-p)\cm_{Y_{m+p+1,n+q}}-iq\cm_{Y_{n+p,m+q+1}}+in\cm_{Y_{m+q,n+p+1}}\nn\\
    &\quad +\frac{1}{2}nq\co_{h_{p+n+1,m+q+1}}-\frac{1}{2}nq\co_{h_{m+q+1,p+n+1}}.
\end{align}
}
}

\paragraph{Super-Poincar\'e algebra.}
Now we check that \eqref{susyicd} can actually reduce to the super-Poincar\'e algebra \cite{Martin:1997ns,2007qft..book.....S} when the parameters are chosen as special values, namely that $f$ takes $n_\mu$, $\bm Y$ takes CKVs $-Y^A_\mn$,  and $\eta,\bar\eta$ take $\l_a,\l^{\dagger\dot a}$.
In other words, we can identify
\bea 
P^\mu\leftrightarrow \mathcal{T}_{n^\mu},\quad M^{\mu\nu}\leftrightarrow \mathcal{M}_{- Y^{\mu\nu}}+\mathcal{T}_{f=-\frac{1}{2}u\nabla\cdot Y^{\mu\nu}},\quad Q_a\leftrightarrow \mathcal{Q}_{\sqrt{2}\lambda^a},\quad \bar{Q}_{\dot a}\leftrightarrow \bar{\mathcal{Q}}_{\sqrt{2}\lambda^{\dagger \dot a}},
\eea 
where for the Lorentz generator, we have included the subtracted part like supertranslation in \eqref{2.79} which is essential for some of the following matchings. 

The pure Poincar\'e algebra will be obtained from the first two lines of \eqref{susyicd}
\begin{align}
  &\begin{cases}
  [\mathcal{T}_{n^\m},\mathcal{T}_{n^\n}]=0,\\
  [\mathcal{M}_{-Y^\mn},\mathcal{M}_{-Y^\rs}]=i\mathcal{M}_{-(-[Y^\mn,Y^\rs])},\\ 
  [\mathcal{T}_{n^\m},\mathcal{M}_{-Y^{\rs}}+\mathcal{T}_{-\frac{1}{2}u\nabla\cdot Y^{\rs}}]=i(\eta^{\m\s}\mathcal{T}_{n^\r}-\eta^{\m\r}\mathcal{T}_{n^\s}),
  \end{cases}\\
  \Rightarrow\quad&
  \begin{cases}
  [P^\m,P^\n]=0,\\ [P^\m,M^\rs]=i\left(\eta^{\m\s}P^\r-(\r\leftrightarrow\s)\right),\\
  [M^\mn,M^\rs]=i\left(\eta^{\m\r}M^{\n\s}-(\m\leftrightarrow\n)\right)-(\r\leftrightarrow\s),
  \end{cases}\label{poincarealgebra}
\end{align} 
where we need the following two identities
\begin{align}
  &\frac{1}{2}n^\m\nabla\cdot Y^\rs-Y^\rs_A\p^An^\m=\eta^{\r\m}n^\s-\eta^{\s\m}n^\r,\\ 
  &[Y_{\mu\nu}, Y_{\rho\sigma}]=\eta_{\nu\rho} Y_{\mu\sigma} + \eta_{\mu\sigma}Y_{\nu\rho} - \eta_{\mu\rho} Y_{\nu\sigma} - \eta_{\nu\sigma} Y_{\mu\rho}.
\end{align}
For the Lorentz algebra, we also have the following matching equation 
\begin{align}
    &[\mathcal{M}_{-Y_{\mu\nu}}+\mathcal{T}_{-\frac{1}{2}u\nabla\cdot Y_{\mu\nu}},\mathcal{M}_{-Y_{\rho\sigma}}+\mathcal{T}_{-\frac{1}{2}u\nabla\cdot Y_{\rs}}]=i\mathcal{M}_{[Y_{\mu\nu}, Y_{\rho\sigma}]}+\frac{i}{2}\mathcal{T}_{u\nabla\cdot [Y_{\mu\nu}, Y_{\rho\sigma}]}
\end{align}
with the part like supertranslation added. For other commutators involving the Lorentz generator, we can only use $\mathcal{M}_{- Y^{\mu\nu}}+\mathcal{T}_{f=-\frac{1}{2}u\nabla\cdot Y^{\mu\nu}}$ but not just $\mathcal{M}_{- Y^{\mu\nu}}$.

Now we concentrate on the commutators involving the superflux. It is easy to find $[\ct_{n^\mu},\cq_{\l_a}]=0$ which implies
\be 
{[P_\mu, Q_a]=0.}\label{PmuQ}
\ee  The next equation that one needs to verify is
\begin{align}
 [\cm_{-Y_\mn}+\ct_{-\frac{1}{2}u\nabla\cdot Y_\mn},\cq_\eta]&=-i\cq_{Y_\mn^A\nabla_A\eta+\frac{1}{8}o(y_\mn,\by_\mn)\eta-\frac{1}{4}\nabla\cdot Y_\mn\eta}\label{3.52}
\end{align}
will reduce to
\begin{align}
    [M^\mn,Q_a]=-(S^\mn_L)_a{}^bQ_b,\label{3.53}
\end{align}
where
\begin{align}
  (S^\mn_L)_a{}^b\equiv\frac{i}{4}(\s^{\mn})_a{}^b=\frac{i}{4}(\s^\m\bar\s^\n-\s^\n\bar\s^\m)_a{}^b.
\end{align}
We need the following identities\footnote{The $y_{\mu\nu}$ and $\bar y_{\mu\nu}$ are defined as $y_{\mu\nu}=Y^A_{\mu\nu}\zeta_A,\ \bar y_{\mu\nu}=Y^A_{\mu\nu}\bar\zeta_A$ (see Appendix \ref{twistor}).}
\begin{subequations}
  \begin{align}
    &\nabla\cdot Y_\mn=\bar n_{\mu} n_{\nu}-\bar n_{\nu} n_{\mu},\\
    &Y^A\nabla_A\l_a=\frac{1}{4}(\by \nabla_A\z^A-y \nabla_A\bar\z^A)\l_a+\frac{1}{2}y\k_a,\\
    &o(y_\mn,\by_\mn)=4y_{[\m}\by_{\n]}-2(\by_\mn \nabla_A\z^A-y_\mn \nabla_A\bar\z^A)
\end{align}
\end{subequations}
to compute the parameter in \eqref{3.52} (omitting an overall $-i$)
\begin{align}
    Y^A_\mn\nabla_A\l_a+\frac{1}{8}o(y_\mn,\by_\mn)\l_a-\frac{1}{4}\nabla\cdot Y_\mn\l_a={ \frac{1}{4}(\s_{\mn})_a{}^b\l_b}
\end{align}
where we have used the identities in \eqref{A14}. 

Now we consider the commutator between superfluxes which will reduce to
\begin{align}
  \{\cq_{\l_a},\bar\cq_{\l^\dagger_{\dot a}}\}=\ct_{\l_a\l^\dagger_{\dot a}}=-\ct_{n_{a\dot a}}=-\ct_{\s^\m_{a\dot a}n_\mu}.
\end{align}
This result is exactly the same as the usual super-Poincar\'e algebra
\begin{align}
    \{Q_a,\bar Q_{\dot a}\}=-2\s^\m_{a\dot a}P_\m.\label{QQP}
\end{align}

\paragraph{Super-Poincar\'e algebra with helicity flux operator.} In the previous discussion, we have stripped off the helicity flux operator 
 which will not appear in the super-Poincar\'e algebra. However, the helicity flux itself will still interact with the superflux although not with Poincar\'e generators. We have 
\begin{align}
  [\co_{h=1},\cq_{\l_a}]=-\cq_{\l_a}\quad\Rightarrow \quad { [H,Q_a]=-Q_a.}\label{OQa}
\end{align} { We have defined $H$ as the global part of the helicity flux operator
\be 
H=\mathcal{O}_{h=1}.
\ee 
All the other commutators involving $H$ are vanishing
\bea 
[P_\mu,H]=[M_{\mu\nu},H]=0.\label{PHMH}
\eea 
 The commutators \eqref{poincarealgebra},\eqref{PmuQ},\eqref{3.53},\eqref{QQP},\eqref{OQa},\eqref{PHMH} and their conjugates form a finite global super-Poincar\'e algebra with helicity operator. Rather interesting, this algebra is isomorphic to the $R$-extended super-Poincar\'e algebra. Indeed, the $R$ symmetry acts on the supercharges as
\bea 
[R, Q_a]=-Q_a,\quad [R,\bar Q_{\dot a}]=\bar Q_{\dot a}.\label{3.88}
\eea Moreover, $H$ indeed commutes with Poincar\'e generators as $R$ symmetry generator does. As a matter of fact, both of them  generate $U(1)$ symmetry. The $U(1)$ associated with $H$ acts on the chiral spinor while $U(1)_R$ symmetry acts on the superflux. However, this does not imply that the helicity flux operator can be identified with a locally extended $R$ flux operator. We will discuss their differences in the following.}

In the superspace formalism, one can realize $R$ symmetry as a (passive) spinor coordinate transformation and hence as an active transformation of components of the chiral superfield \cite{Martin:1997ns}
\bea 
\mathcal{C}(x,\theta,\bar\theta)=\Phi(x)+\sqrt{2}\theta \chi(x) +\cdots,
\eea where we have omitted the non-dynamical field in the superspace. The $U(1)_R$ symmetry acts on the superspace coordinates $\theta$ as 
\be \theta\to \theta'=e^{i{\beta}}\theta.
\ee The superfield $\mathcal{C}$ with charge $\tt r$ will transform as 
\be 
\mathcal{C}(x,\theta,\bar\theta)\to \mathcal{C}'(x,e^{i{\beta }}\theta,e^{-i{\beta}}\bar\theta)=e^{i{\tt r}\beta}\mathcal{C}(x,\theta,\bar\theta),
\ee 
whose components transform as
\begin{align}
  \Phi\to e^{i{\tt r}\beta}\Phi,\quad \chi\to e^{i(\tt r-1){\beta}}\chi.
\end{align}
If we set $\tt r=0$ and then the field $\Phi$ and $\chi$ will transform as 
\bea 
\Phi(x)\to \Phi'(x)=\Phi(x),\quad \chi(x)\to \chi'(x)=e^{-i\beta}\chi(x).
\eea Note that it transforms the fermionic field by a phase while keeping the scalar field invariant. This agrees with the transformation law  \eqref{chiralwz} acted by the helicity flux operator with 
\be 
h=\beta.
\ee  However, we note that the choice of $\tt r=0$ does not preserve the action due to the Yukawa term. To preserve the action, the $R$ charge should be chosen as $2/3$.

In summary, we can add a helicity flux operator $H$ to the super-Poincar\'e algebra and the resulting algebra is  isomorphic to $R$-extended super-Poincar\'e algebra. Using the terminology in our previous paper, the helicity flux operator generates a  superduality transformation. One cannot identify it with a local $R$ symmetry transformation due to the mismatch of the $R$ charge. 

{\paragraph{Charge flux for complex scalar in $R$-extended super-Poincar\'e algebra.} As stated above, we do not include the flux $\mathcal{E}^{\rm b}_e$ defined in \eqref{3.49} for complex scalar since it will not emerge naturally.  However, once reducing to $R$-extended super-Poincar\'e algebra, it is necessary to note that $\ce^{\rm b}_{e=1/2}$  should be added to the $R$ symmetry generator. One can easily check the above statement by computing
\begin{align}
  [\ce^{\rm b}_{e=1/2},\ct_{n^\mu}]=[\ce^{\rm b}_{e=1/2},\cm_{-Y^\mn}]=0,
\end{align}
and
\begin{align}
  [\ce^{\rm b}_{e=1/2},\cq_{\l_a}]=-\cq_{\l_a},\qquad
  [\ce^{\rm b}_{e=1/2},\bar\cq_{\l^\dagger_{\dot a}}]=\bar\cq_{\l^\dagger_{\dot a}}.
\end{align}
It is reasonable to include both bosonic part $\ce^{\rm b}_{e=1/2}$ and fermionic part $H$ in the $R$-extended super-Poincar\'e algebra. Therefore, we should define a unified $R$ flux
\begin{align}
    \cR=\frac{2}{3}\ce^{\rm b}_{e=1/2}+\frac{1}{3}H\label{cr}
\end{align}
such that $\cR,\ct_{n^\mu},\cm_{-Y^\mn},\cq_{\l_a}$ and $\bar\cq_{\l^\dagger_{\dot a}}$ completely realize the $R$-extended super-Poincar\'e algebra in the Wess-Zumino theory. This $R$ flux acts on fields as expected
\begin{align}
    [\mathcal{R},F(u,\Omega)]=-\frac13F(u,\Omega),\quad [\mathcal{R},\Sigma(u,\Omega)]=\frac23\Sigma(u,\Omega).
\end{align}

\paragraph{Generalized $R$ flux and its commutators.} Note that \eqref{cr} is the total flux which can be derived from the $R$ symmetry of the Wess-Zumino model \cite{Martin:1997ns,Shifman:2012zz}
\begin{align}
  \Phi(x)\to e^{2i{\beta}/3}\Phi(x), \qquad  \chi(x)\to e^{-i{\beta}/3}\chi(x),\label{rsym}
\end{align}
using Noether's procedure. The conserved current is \cite{Shifman:2012zz}
\begin{align}
  J_R^\m&=\frac{\p \cl}{\p\p_\m\Phi}\delta\Phi+\frac{\p \cl}{\p\p_\m\bar\Phi}\delta\bar\Phi+\frac{\p \cl}{\p\p_\m\chi}\delta\chi\nn\\
  &=-\frac{2i}{3}(\Phi\p^\m\bar\Phi-\bar\Phi\p^\m\Phi)+\frac{1}{3}\bar\chi\bar\s^\m\chi,
\end{align}
which gives the following flux at $\ci^+$
\begin{align}
  \cf_R=r^{2}\int dud\Omega\,m_\m J_R^\m=\frac{1}{3}\int dud\Omega\,\Big[2i(\Sigma\dot{\bar{\Sigma}}-\bar\Sigma\dot{{\Sigma}})+2\bar FF\Big]
\end{align}
 which is exactly  $\cR$. It is natural to define a generalized $R$ flux
\begin{align}
  \cR_{\tt r}=\co_h+\ce^{\rm b}_e=\int d u d \Omega\, {\tt r}(\Omega) [2i(:\Sigma\dot{\bar{\Sigma}}-\bar\Sigma\dot{{\Sigma}}:)+2:\bar FF:],
\end{align}
which acts on fields as
\begin{align}
  &[\cR_{\tt r},F(u,\Omega)]=-{{\tt r}(\Omega)}F(u,\Omega),\qquad [\cR_{\tt r},\bar F(u,\Omega)]={{\tt r}(\Omega)}\bar F(u,\Omega),\\
  &[\cR_{\tt r},\Sigma(u,\Omega)]=2{\tt r}(\Omega)\Sigma(u,\Omega),\qquad [\cR_{\tt r},\bar\Sigma(u,\Omega)]=-2{\tt r}(\Omega)\bar\Sigma(u,\Omega).
\end{align}
We can also compute the algebra involving this $R$ flux operator. For simplicity, we just consider to extend the largest Lie algebra, namely \eqref{lwzal}, and only display the commutators involving $\cR_{\tt r}$ 
\begin{subequations}
  \begin{align}
    &[\mathcal{T}_f,\cR_{\tt r}]=0,\qquad [\mathcal{M}_{\bm Y},\cR_{\tt r}]=i\mathcal{R}_{\bm Y({\tt r})},\qquad [\mathcal{R}_{{\tt r}_1},\mathcal{R}_{{\tt r}_2}]=0,\\
    &[\cR_{\tt r},\cq_\eta]=-3\cq_{{\tt r}\eta},\qquad [\cR_{\tt r},\bar\cq_{\bar\eta}]=3\bar\cq_{{\tt r}\bar\eta}.\label{oq3}
  \end{align}
\end{subequations}
The factor 3 in the \eqref{oq3} exactly makes {it} matching with \eqref{3.88} after taking ${\tt r}=1/3$. Except the above, we have already known that there is only fermionic helicity flux appears in the right-hand sides of the following two commutators
\begin{subequations}
\begin{align}
&[\mathcal{M}_{\bm Y},\mathcal{M}_{\bm Z}]=i\mathcal{M}_{[\bm Y, \bm Z]}-\frac{i}{2}\co_{o(\bm Y, \bm Z)},\\
&[\cq_{\eta_1},\bar\cq_{\bar\eta_2}]=-\ct_{u^2\mu_1\bar\m_2+u(\m_1\bar\nu_2+\n_1\bar\mu_2)+\n_1\bar\n_2}-\frac{i}{2}\co_{\m_1\bar\n_2+\bar\m_2\n_1}.
\end{align}
\end{subequations}
Thus it is necessary to consider the commutator between $\co_h$ and $\cR_{\tt r}$
\begin{align}
    [\co_h,\cR_{\tt r}]=0.
\end{align}
The charge flux of complex scalar will not appear separately in this algebra.
}

\section{Discussion and conclusion}\label{dis}
In this work, we have extended Carrollian diffeomorphism in the Dirac and Wess-Zumino theory. The main results are summarized in \eqref{spinoralgebrap} and \eqref{susyicd} respectively. 
In the Dirac theory, we add a helicity flux operator to close the algebra. Similar to the bosonic cases, the helicity flux operator generates the superduality transformation, which is a local chiral symmetry transformation of the boundary fermionic field. In contrast with the electromagnetic/gravitational helicity flux operator, the test function $h$ in the newly found helicity flux operator $\mathcal{O}_h$ could be time-dependent. Note that $h$ is time-independent in the bosonic theory to remove the non-local terms in the algebra. However, there are no such annoying  terms in the Dirac theory.  As a result, the corresponding infinite-dimensional algebra has two central charges, extending the intertwined Carrollian diffeomorphism  found in bosonic theories. 
We also combine  Carrollian diffeomorphism  with supersymmetry successfully in the Wess-Zumino theory. Interestingly, since the supertranslation function $f$ is time-dependent, the Grassmann field $\eta$ in the superflux can also be lifted to be time-dependent. As a consequence, the $\mathcal{N}=1$ supersymmetric extension of Carrollian diffeomorphism can have three central charges. Moreover, the commutator of the superfluxes leads  to the superposition of the energy flux and helicity flux operators which is reflected in the left equation of \eqref{qq}. When all the functions are time-independent, our result is consistent with the supersymmetric BMS algebra in the literature.

There are many open issues in this direction. 
\begin{itemize}
\item \textbf{Violation of Jacobi identity.}  There is a long history of studying the failures of Jacobi identities in the literature \cite{PhysRev.149.1268,Johnson:1966se,Brandt:1968zza,Lipkin:1969ck,Jackiw:1984rd,Jo:1985jf,Levy:1986yr,Banerjee:1990qv}. In particular, it was shown that this failure may relate to the axial anomaly according to \cite{Levy:1986yr} in the context of current algebra. Recently, an example in celestial CFT states that the violation of Jacobi identities originates to the non-commutativity of the soft limits \cite{Mago:2021wje,Ren:2022sws,Ball:2024oqa} where the authors argued that this violation prevents us from identifying the soft current as a symmetry generator to be included in the symmetry algebra, but has no ill effect for the OPE.

In our case, the violation of Jacobi identity comes from the central extensions which actually do not change the action of the flux operators on the boundary field. Its role is to match the vacuum expectation values of the two sides. Of course, a central extension of a Lie algebra usually should satisfy the Jacobi identity which for instance determines the form of the central term of Virasoro algebra uniquely. According to \cite{Silva1999GeometricMF}, an algebra equipped with a bilinear antisymmetric bracket like a Lie algebra is called almost Lie algebra whose bracket does not necessarily satisfy Jacobi identity. Our flux algebra \eqref{spinoralgebrap} is an almost Lie algebra, but unfortunately we do not find much study on it. 

\paragraph{Various relaxations of Jacobi identity.} We consider several examples:
\begin{enumerate}
    \item When deforming the Virasoro algebra, the so-called Hom-Lie algebra was introduced \cite{Aizawa:1990hx,2005math.....12526H,Hartwig_2006} where the Hom-Jacobi identity with respect to a bracket-preserving linear map on the Lie algebra is satisfied
    \begin{align}
        [g(x),[y,z]]+[g(y),[z,x]]+[g(z),[x,y]]=0,\quad \text{for}\ \forall x,y,z\in\fg,\  \text{with}\  g:\fg\to\fg.
    \end{align}
    When $g(x)=x$, this reduces to the usual Lie algebra. It seems impossible to find such a map over our flux algebra which can cure the problem of the Jacobi identity violation. 
        \item Another popular one is the strong homotopy (sh) Lie algebra or $L_\infty$ algebra \cite{Lada:1992wc,Schlessinger1985TheLA,10.1007/BFb0101184} where the Jacobi identity holds only up to compatible higher homotopies (or called BRST exact terms). 
   In our treatment, total derivatives with respect to $u$ are discarded and thus do not affect the Jacobi identity. The Jacobi violations \eqref{jacvio} have the structure $\int dud\Om\,Y^A\nabla_A(\cdots)$ where the dots represent other parameters, and can not be formed into a total derivative. Therefore, the sh Lie algebra can not be the solution to our problem.
    
    \item If violating Jacobi identity, one can require
    \begin{align}
        J[x,y,[x,z]]=[J[x,y,z],x]\quad \text{for}\ \forall x,y,z\in\fg,
    \end{align}
    so that it form a Malcev algebra \cite{Malcev,Sagle}.  In our case, the Jacobiator $J[x,y,z]$ is a constant and thus $[J[x,y,z],x]=0$. However, we find $J[\ct_{f_1},\cm_{\bm Y},[\ct_{f_1},\ct_{f_2}]]\ne0$, so it is not a Malcev algebra. Furthermore, $J[\cm_{\bm Y},\co_{h_1 },[\cm_{\bm Y},\co_{h_2}]]\ne0$ implies that the flux operators $\cm_{\bm Y}$ and $\co_h$ do not form a Malcev algebra. 
\end{enumerate}

\item \textbf{Non-closure of the Lie transport of the spinor field around a loop.} As is shown in \eqref{anomalous}, the Lie transport of the spinor field around a loop is not closed  which is in contrast with the closure of the Lie transport of the tensor field. In our paper, the non-closure property is the origin of the anomalous helicity flux operator in the commutator of superrotation generators for the spinor field. The Lie derivative of the spinor field in a Riemann manifold calls for spin structure of the manifold \cite{26273c3a-85bd-3898-9eca-8ef65e14882f}. There should be a deep connection between the helicity flux operator and the geometry of the manifold. 
    \item \textbf{Chiral anomaly and helicity flux operator.} In massless QED, there is a famous chiral anomaly stating that the divergence of the chiral current $j^\mu_5$ is non-zero at the quantum level
    \be 
\partial_\mu j^\mu_5=-\frac{e^2}{16\pi^2}\epsilon^{\mu\nu\rho\sigma}f_{\mu\nu}f_{\rho\sigma},\label{anomaly}
\ee where $f_{\mu\nu}$ is the   strength tensor of the  electromagnetic field and $e$ is the electric charge. 
In this paper, we have shown that the chiral current can be used to define the helicity flux operator $\mathcal{O}_h$ at future null infinity. Interestingly, 
\bea 
\mathcal{O}_{h=1}=\int (d^3x)_\mu j^\mu_5=\# \ \text{positive helicity}-\#\ \text{negative helicity}
\eea is exactly the Atiyah-Singer index of the Dirac operator \cite{Nakahara:2003nw}. On the other hand, the right hand side can be formed to a total derivative and a topological term since
\begin{align}
    f\wedge f=-\frac{1}{4}\epsilon^{\mu\nu\rho\sigma}f_{\mu\nu}f_{\rho\sigma}d^4x=-\frac{1}{2}\p_\m (\epsilon^{\mu\nu\rho\sigma}f_{\nu\r}a_{\sigma})d^4x.
\end{align}
More important, it can also give rise to the helicity flux $\mathcal{O}^{\rm v}_{h=1}$ of the Maxwell theory \cite{Liu:2024rvz}
\bea 
-\frac{e^2}{8\pi^2} \int_{\ci^+} (d^3x)_\mu\,\epsilon^{\mu\nu\rho\sigma}f_{\nu\r}a_{\sigma}=-\frac{e^2}{4\pi^2}\int du d\Omega A_A \dot{A}_B \epsilon^{AB}\equiv\frac{e^2}{4\pi^2}\mathcal{O}^{\rm v}_{h=1}.
\eea 
Therefore,  the integrated anomaly equation at $\mathcal{I}^+$ may be regarded as a balance equation between helicity fluxes of spinor and vector fields 
\bea 
\text{Helicity flux of Dirac field}=\text{Helicity flux of Maxwell field}.
\eea In figure \ref{triangle}, we summarize the previous discussion as a triangle relation.  
It would be interesting to explore the relation among chiral anomaly, helicity flux operator and topological term in the future.
\begin{figure}
    \begin{center}
          \begin{tikzpicture}
    \node[draw, rounded corners] (0) {$\begin{matrix}
      \displaystyle \p_\m j^\m_5=-\frac{e^2}{16\pi^2}\epsilon^{\mu\nu\rho\sigma}f_{\mu\nu}f_{\rho\sigma}\\
      \text{chiral anomaly equals the second Chern character}
    \end{matrix}$};
    \coordinate (1) at (0,-3);
    \node[draw, rounded corners, left=1cm of 1] (2) {$\begin{matrix}
      \displaystyle \co_{h=1}\sim \int_{\ci^+}(d^3x)_\m j^\m_5\\
      \text{helicity flux for massless Dirac spinor}
    \end{matrix}$};
    \node[draw, rounded corners, right=1cm of 1] (3) {$
    \begin{matrix}
      \displaystyle \co^{\rm v}_{h=1}\sim \int_{\ci^+}I,\ \ dI=f\wedge f\\
      \text{helicity flux for Maxwell field}
    \end{matrix}$};
    \draw[black!30,line width=2pt] (0) -- (2);
    \draw[black!30,line width=2pt] (0) -- (3);
    \draw[black!30,dashed,line width=2pt] (2) -- (3);
  \end{tikzpicture}
    \end{center}
    \caption{The triangle relation among the helicity flux operators of the Dirac field and the Maxwell field as well as the topological Chern character.}
    \label{triangle}
\end{figure}
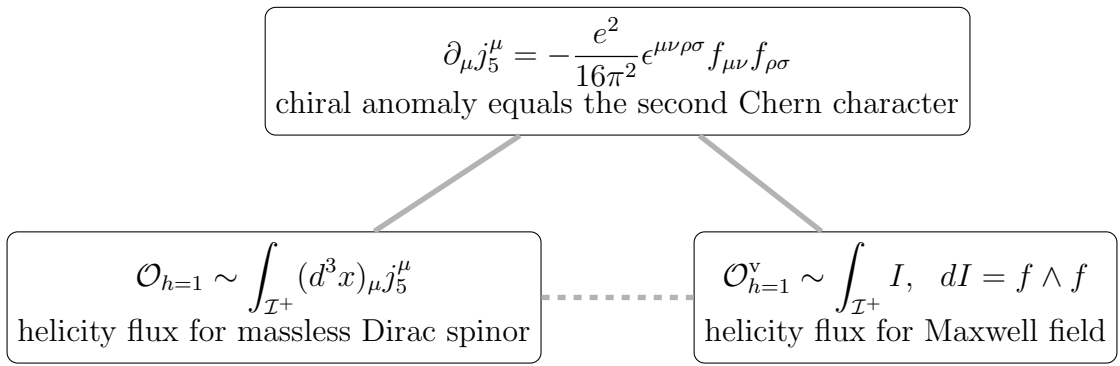

\item \textbf{Carrollian supergeometry.} { In a concrete approach \cite{Rogers2007}, one can equip the superspace with DeWitt topology \cite{DeWitt:2012mdz} to define a supermanifold. It would be interesting to explore the supergeometry for a Carrollian manifold, in the methods of either bulk reduction (as a hypersurface) or focusing on the Carrollian manifold its own.}
\item \textbf{Carrollian amplitude involving spinors.} The  bulk reduction for the spinor field near $\mathcal{I}^+$ in this work can be used to study Carrollian amplitude \cite{Liu:2024llk, Liu:2022mne, Donnay:2022wvx, Salzer:2023jqv,Nguyen:2023miw, Mason:2023mti,Liu:2024nfc,Adamo:2024mqn,Alday:2024yyj,Ruzziconi:2024zkr,Li:2024kbo} when there are fermions and supersymmetry. {In principle, we can use the same technology to study the possible modification of the algebras systematically in the presence of interactions.} We will investigate this problem in the future. 
\end{itemize}

\vspace{10pt}
{\noindent \bf Acknowledgments.} 
The work of J.L. was supported by NSFC Grant No. 12005069.  The work of W.-B. Liu is supported by ``the Fundamental Research Funds for the Central Universities'' with No. YCJJ20242112.

\appendix
\section{Twistor identities}\label{twistor}
In this section, we will collect the twistor identities which are useful in the context. 
\begin{enumerate}
    \item Contractions of two twistors: 
    \bs\label{contrac}\begin{align}
       & \lambda^a\lambda_a=\kappa^a\kappa_a=\lambda^\dagger_{\dot a}\lambda^{\dagger\dot a}=\kappa^\dagger_{\dot a}\kappa^{\dagger\dot a}=0,\\
        &\kappa^a\lambda_a=-\lambda^a\kappa_a=-2,\quad \lambda^{\dagger\dot a}\kappa^\dagger_{\dot a}=-\kappa^{\dagger\dot a}\lambda^\dagger_{\dot a}=2.
    \end{align}
    \es We can omit the indices and rewrite the identities as follows 
    \bs\begin{align}
        &\lambda^2=\kappa^2=\left(\lambda^{\dagger}\right)^2=\left(\kappa^{\dagger}\right)^2=0,\\
        &\kappa\lambda=-\lambda\kappa=-2,\quad \kappa^\dagger\lambda^{\dagger}=-\lambda^\dagger\kappa^\dagger=2.
    \end{align}\es 
    \item Vectors $n^\mu,\bar n^\mu, m^\mu,\bar m^\mu, Y^\mu_A$ can be expressed as combinations of twistors
    \bs\label{comb}\begin{align}
        &n_{a\dot a}=-\lambda_a\lambda^{\dagger}_{\dot a},\\
        &\bar n_{a\dot a}=\kappa_a\kappa^\dagger_{\dot a},\\
        &m_{a\dot a}=\frac{1}{2}(\kappa^\dagger_{\dot a}\kappa_a-\lambda^\dagger_{\dot a}\lambda_a),\\
        &\bar m_{a\dot a}=-\frac{1}{2}(\lambda^\dagger_{\dot a}\lambda_a+\kappa^\dagger_{\dot a}\kappa_a),\\
        &Y^A_{a\dot a}=\frac{1}{2}(\lambda_a\kappa^\dagger_{\dot a}\bar\zeta^A+\kappa_a\lambda^\dagger_{\dot a}\zeta^A).
    \end{align}\es In the last equation, we introduced the vectors on the celestial sphere $S^2$
\be 
\zeta^A=\left(\begin{array}{c}1\\ -\frac{i}{\sin\theta}\end{array}\right),\quad \bar{\zeta}^{A}=\left(\begin{array}{c}1\\ \frac{i}{\sin\theta}\end{array}\right)\label{zeta}
\ee
that satisfy the following identities \bea 
&&\zeta^A\zeta_A=0,\quad \zeta^A\bar{\zeta}_A=2,\quad \gamma^{AB}=\frac{1}{2}(\zeta^A\bar{\zeta}^B+\zeta^B\bar{\zeta}^A),\quad \epsilon^{AB}=-\frac{i}{2}(\zeta^A\bar{\zeta}^B-\zeta^B\bar{\zeta}^A).\nn\\ \label{zetaidentity}
\eea 
Any vector $V^A$ on the sphere can be decomposed as 
\bea 
V^A=\frac{1}{2}(\zeta^A\bar v+\bar{\zeta}^Av)
\eea with 
\be 
v=V^A\zeta_A,\quad \bar v=V^A\bar\zeta_A.
\ee 
The covariant derivative of the vectors $\zeta^A,\bar{\zeta}^A$ are
\bs\begin{align}
\zeta^B\nabla_B\zeta_A&=\nabla_B\zeta^B \zeta_A,\\
\zeta^B\nabla_B\bar{\zeta}_A&=-\nabla_B\zeta^B \bar{\zeta}_A,\\
\bar{\zeta}^B\nabla_B\zeta_A&=-\nabla_B\bar{\zeta}^B \zeta_A,\\
\bar{\zeta}^B\nabla_B\bar{\zeta}_A&=\nabla_B\bar{\zeta}^B \bar{\zeta}_A,\\
\zeta^B\nabla_A\bar{\zeta}_B&=-\bar{\zeta}^B\nabla_A\zeta_B=\zeta_A\nabla_B\bar{\zeta}^B-\bar{\zeta}_A\nabla_B{\zeta}^B.
\end{align}\es  We also find the following two identities 
\be 
\nabla_A\zeta^A=\nabla_A\bar{\zeta}^A=\cot\theta.
\ee  
\bea   
\bar{\zeta}_C\nabla_A\zeta_B-\zeta_B\nabla_A\bar{\zeta}_C+\zeta_C\nabla_A\bar{\zeta}_B-\bar{\zeta}_B\nabla_A\zeta_C=2i\epsilon_{BC}\bar{\zeta}^D\nabla_A\zeta_D.
\eea  
\item Contractions with Pauli matrices:
\bs\begin{align}
   & \lambda^a\sigma^\mu_{a\dot a}\lambda^{\dagger\dot a}=2n^\mu,\quad \lambda^a\sigma^\mu_{a\dot a}\kappa^{\dagger\dot a}=-2y^\mu,\quad \kappa^a\sigma^\mu_{a\dot a}\lambda^{\dagger\dot a}=-2\bar y^\mu,\quad \kappa^a\sigma^\mu_{a\dot a}\kappa^{\dagger\dot a}=-2\bar n^\mu,\\
   &\lambda^{\dagger}_{\dot a}\bar\sigma^{\mu\dot aa}\lambda_a=2n^\mu,\quad \lambda^\dagger_{\dot a}\bar\sigma^{\mu\dot aa}\kappa_a=-2\bar y^\mu,\quad \kappa^{\dagger \dot a}\bar\sigma^{\mu\dot aa}\lambda_a=-2y^\mu,\quad \kappa^{\dagger}_{\dot a}\bar\sigma^{\mu\dot a a}\kappa_a=-2\bar n^\mu.
\end{align}\es Note that we have defined 
\bea 
y^\mu=Y^{\mu A}\zeta_A,\quad \bar y^\mu=Y^{\mu A}\bar\zeta_A.
\eea Both of them can be transformed to the matrix forms
\bea 
y_{a\dot a}=y^\mu\sigma_{\mu a \dot a}=\lambda_a\kappa^\dagger_{\dot a},\quad \bar y_{a\dot a}=\bar y^\mu\sigma_{\mu a\dot a}=\kappa_a\lambda^\dagger_{\dot a}.
\eea The above equations can also be expressed as \bs\label{A14}
\begin{align}
    \sigma^\mu_{a\dot a}\lambda^{\dagger\dot a}&=\bar y^\mu\lambda_a+n^\mu\kappa_a,\\
    \sigma^\mu_{a\dot a}\kappa^{\dagger\dot a}&=\bar n^\mu\lambda_a-y^\mu\kappa_a,\\
    \bar\sigma^{\mu\dot a a}\lambda_a&=-y^\mu\lambda^{\dagger\dot a}-n^\mu\kappa^{\dagger\dot a},\\
    \bar\sigma^{\mu\dot a a}\kappa_a&=-\bar n^\mu\lambda^{\dagger\dot a}+\bar y^\mu\kappa^{\dagger\dot a}.
\end{align}\es 
The conformal Killing vectors $Y_{\mu\nu}^A=Y_\mu^A n_\nu-Y_\nu^A n_\mu$ can be contracted   \bs\begin{align}
   Y^A_{\mu\nu}{}^{\dot a}_{\ \dot b}&= Y^A_{\mu\nu}\left(\bar\sigma^\mu\sigma^\nu\right)^{\dot a}_{\ \dot b}=2\lambda^{\dagger\dot a}\lambda^\dagger_{\dot b}\zeta^A,\\
   Y^A_{\mu\nu a}{}^b&=Y^A_{\mu\nu}\left(\sigma^\mu\bar\sigma^\nu\right)_a^{\ b}=-2\lambda_a\lambda^b\bar\zeta^A,\\
  \bar{Y}^A_{\mu\nu}{}^{\dot a}_{\ \dot b}&=\bar Y^A_{\mu\nu}\left(\bar\sigma^\mu\sigma^\nu\right)^{\dot a}_{\ \dot b}=2\kappa^\dagger_{\dot a}\kappa^{\dagger\dot b}\bar\zeta^A,\\
  \bar Y^A_{\mu\nu a}{}^b&=\bar Y^A_{\mu\nu}\left(\sigma^\mu\bar\sigma^\nu\right)_a^{\ b}=-2\kappa_a\kappa^b\zeta^A,
\end{align}\es 
where we have similarly defined 
\be 
\bar Y_{\mu\nu}^A=Y_\mu^A\bar n_\nu-Y_\nu^A n_\mu.
\ee There is another tensor which is useful \be n_{\mu\nu}=n_\mu\bar n_\nu-n_\nu\bar n_\mu,
\ee which can be contracted with $\sigma^{\mu\nu}$ and $\bar\sigma^{\mu\nu}$ to obtain 
\bea 
n_{\mu\nu}(\sigma^{\mu\nu}){}_a^{\ b}=4(\lambda_a\kappa^b+\kappa_a\lambda^b),\quad n_{\mu\nu}(\bar\sigma^{\mu\nu}){}^{\dot a}_{\ \dot b}=-4(\lambda^{\dagger\dot a}\kappa^{\dagger}_{\dot b}+\kappa^{\dagger\dot a}\lambda^{\dagger}_{\dot b}).
\eea 
\item Covariant derivatives of twistors:
\bs\begin{align}
&\zeta^A\nabla_A\lambda_a=\frac{1}{2}\nabla_A\zeta^A \lambda_a,\quad \bar\zeta^A\nabla_A\lambda_a=-\frac{1}{2}\nabla_A\bar{\zeta}^A\lambda_a+\kappa_a,\\&\zeta^A\nabla_A{\lambda}^\dagger_{\dot a}=-\frac{1}{2}\nabla_A\zeta^A {\lambda}^{\dagger}_{\dot a}+{\kappa}^\dagger_{\dot a},\quad \bar\zeta^A\nabla_A{\lambda}^\dagger_{\dot a}=\frac{1}{2}\nabla_A\bar\zeta^A {\lambda}^\dagger_{\dot a},\\
&\zeta^A\nabla_A\kappa_a=-\lambda_a-\frac{1}{2}\nabla_A\zeta^A\kappa_a,\quad \bar{\zeta}^A\nabla_A\kappa_a=\frac{1}{2}\nabla_A\bar{\zeta}^A\kappa_a,\\
& \zeta^A\nabla_A\kappa^\dagger_{\dot a}=\frac{1}{2}\nabla_A\zeta^A\kappa^\dagger_{\dot a},\quad \bar{\zeta}^A\nabla_A{\kappa}^\dagger_{\dot a}=-\lambda^\dagger_{\dot a}-\frac{1}{2}\nabla_A\bar\zeta^A\kappa^\dagger_{\dot a}.
\end{align}\es 
\end{enumerate}
\bibliography{refs}
\end{document}